\documentclass[fleqn,usenatbib]{mnras}

\usepackage{newtxtext,newtxmath,comment}
\usepackage[normalem]{ulem}

\usepackage[T1]{fontenc}
\usepackage{ae,aecompl}

\usepackage{graphicx}	
\usepackage{amsmath}	

\title[Inferring Warm Dark Matter Masses]{Inferring Warm Dark Matter Masses with Deep Learning}

\author[J. C. Rose et al.]{
Jonah C. Rose$^{1}$\thanks{E-mail: j.rose@ufl.edu},
Paul Torrey$^{1}$,
Francisco Villaescusa-Navarro$^{2,3}$,
Mark Vogelsberger$^{4,5}$,
\newauthor
Stephanie O'Neil$^{4}$,
Mikhail~V. Medvedev$^{6,7}$,
Ryan Low$^{6}$,
Rakshak Adhikari$^{6}$,
\newauthor
and Daniel Angl\'es-Alc\'{a}zar$^{2,8}$
\vspace{0.3cm}\\
$^{1}$Department of Astronomy, University of Florida, Gainesville, FL 32611, USA \\
$^{2}$Center for Computational Astrophysics, Flatiron Institute, 162 5th Avenue, New York, NY, 10010, USA \\
$^{3}$Department of Astrophysical Sciences, Princeton University, Peyton Hall, Princeton NJ 08544, USA \\
$^{4}$Kavli Institute for Astrophysics and Space Research, Massachusetts Institute of Technology, 70 Vassar St., Cambridge, MA 02139, USA \\
$^{5}$ The NSF AI Institute for Artificial Intelligence and Fundamental Interactions, Massachusetts Institute of Technology, Cambridge, MA 02139, USA \\
$^{6}$Department of Physics and Astronomy, University of Kansas, Lawrence, KS 66045, USA \\
$^{7}$Laboratory for Nuclear Science, Massachusetts Institute of Technology, Cambridge, MA 02139, USA \\
$^{8}$Department of Physics, University of Connecticut, 196 Auditorium Road, U-3046, Storrs, CT, 06269, USA
}

\date{Accepted XXX. Received YYY; in original form ZZZ}

\pubyear{2023}

\begin{document}
\label{firstpage}
\pagerange{\pageref{firstpage}--\pageref{lastpage}}
\maketitle

\begin{abstract}
We present a new suite of over 1,500 cosmological N-body simulations with varied Warm Dark Matter (WDM) models ranging from 2.5 to 30 keV.
We use these simulations to train Convolutional Neural Networks (CNNs) to infer WDM particle masses from images of DM field data.
Our fiducial setup can make accurate predictions of the WDM particle mass up to 7.5 keV 
at a 95\% confidence level from small maps that cover an area of $(25~h^{-1}{\rm Mpc})^2$.
We vary the image resolution, simulation resolution, redshift, and cosmology of our fiducial setup to better understand how our model is making predictions.
Using these variations, we find that our models are most dependent on simulation resolution, minimally dependent on image resolution, not systematically dependent on redshift, and robust to varied cosmologies.
We also find that an important feature to distinguish between WDM models is present with a linear size between 100 and 200 h$^{-1}$ kpc.
We compare our fiducial model to one trained on the power spectrum alone and find that our field-level model can make 2$\times$ more precise predictions and can make accurate predictions to 2$\times$ as massive WDM particle masses
when used on the same data.
Overall, we find that the field-level data can be used to accurately differentiate between WDM models and contain more information than is captured by the power spectrum.
This technique can be extended to more complex DM models and opens up new opportunities to explore alternative DM models in a cosmological environment.
\end{abstract}



\section{Introduction}
\label{sec:introduction}

The $\Lambda$ cold dark matter ($\Lambda$CDM) model has done an excellent job at modeling many aspects of our Universe from large-scale structure down to properties of individual galaxies. 
In this model, Dark Matter (DM) is the backbone of cosmic structure formation and is assumed to be made of \textit{cold} and \textit{collisionless} particles (so-called cold dark matter; CDM).
We have multi-pronged evidence that supports a simple CDM model including 
the rotation curves of galaxies \citep{1996Navarro, 2008Blok}, 
the offset of mass and light in colliding systems \citep[e.g. the Bullet cluster;][]{2006Clowe},
the strength of gravitational lensing signatures \citep{2002Dalal, 2005Waerbeke}, 
the existence of baryon acoustic oscillations \citep{2005Eisenstein, 2014Planck},
and more.
Indeed, the widespread acceptance of a simple cold and collisionless DM particle model is justified given the wide range of observations it can explain within the $\Lambda$CDM framework.

However, some tensions exist between CDM predictions and extragalactic observations -- especially at small scales.
Specifically, 
the \textit{core-cusp}~\citep{Moore1994, 1997Blok}, 
\textit{too-big-to-fail}~\citep[TBTF;][]{Boylan-Kolchin2011, 2015Papastergis}, 
\textit{diversity}~\citep{Oman2015}, and 
\textit{missing satellites}~\citep{1999Moore, Klypin1999} problems represent tensions that have been identified where the standard predictions of $\Lambda$CDM may deviate from what is observed in nearby extragalactic systems~\citep{Bullock2017, Tulin2018}.
While simulations that include not only hydrodynamics but also full galaxy formations models have shown that some tensions (e.g., missing satellites and core-cusp) can be alleviated or resolved without altering the CDM model \citep{2014Vogelsberger, 2015Chan, 2018Simpson, 2021Engler}, others (e.g. TBTF and diversity) may yet require tweaks to the CDM model to resolve \citep{2018Lovell, 2020Callingham}.
These tensions, coupled with the fact that no successful direct detection experiments have been conducted to date that provide any definite proof of the nature of DM \citep{2012Chatrchyan, 2013Aad, 2014Aad}, give rise to a credible basis for considering DM models going beyond simple CDM.

Several DM model variations, including self-interacting DM~\citep[SIDM;][]{Spergel2000} and fuzzy DM~\citep{Hu2000}, have been proposed.
These models can solve some of the discrepancies observed with the CDM model while retaining the successes of CDM on large scales \citep{2012Vogelsberger, 2013Rocha, 2016Vogelsberger, 2017Xiaolong, 2017Hui}.
Unfortunately, the non-linear nature of structure formation makes mapping these fundamental DM particle assumptions onto observable galaxy properties challenging.  
The most efficient tool for making detailed predictions for the growth of structure in non-linear regimes is cosmological numerical simulations.

A great effort has been carried out by the community to study various statistics like the power spectrum, the abundance of subhalos, or the density profiles, to identify high signal-to-noise statistics that allow to better discriminate among different DM models ~\citep[for a review see][]{2017Arun}. Another possibility is to make use of machine learning techniques to carry out that task. It has been shown in different scenarios that these methods typically yield the tightest constraints on the considered parameters since they can explore the entire distribution of statistics and identify the one that performs best \citep{2022Villaescusa, 2021Fluri, 2022Makinen}. Thus, in this work, we made use of machine learning techniques to infer the properties of dark matter.

The CAMELS project \citep{CMD, CAMELS, PublicRelease} has used a large suite of simulations with varied cosmology and baryonic physics coupled with machine learning techniques to better understand the impact of coupled processes on structure formation.
The CAMELS simulation suites have been used to show that CNNs can help extract maximal cosmological information from astrophysical features \citep{2021VillaescusaA, 2021Villaescusa}.
In \cite{CAMELS}, the authors found that they could utilize CNNs to marginalize over baryonic processes to accurately predict the underlying cosmology.
So far, the CAMELS simulation suites have been limited to the CDM model.
However, the method can be extended to explore the alternative DM parameter space in the same fashion.
The simplest DM model that lends itself for an initial exploration into the alternative DM parameter space is warm dark matter (WDM) \citep{2000Hogan, 2001Sommer} owing to its single parameter and clear effect on cosmic structure.

WDM ascribes a significant velocity to the early DM field and has been explored as a method to alleviate the missing satellite and core-cusp problems in Nbody simulations of CDM \citep{2001Bode, 2007Gilmore, 2010Vega, 2013Viel, 2014Lovell}.
These higher initial DM velocities act to dampen the small-scale structure in CDM simulations by allowing DM particles to escape from small potential wells in the early universe.
The single parameter in this model is the mass of the DM particle, measured in keV.
The intensity of the astrophysical smoothing depends on the mass of the DM particle and produces more suppression for lighter (warmer) models. 

Previous works have conducted WDM simulations to date to understand whether WDM can accurately describe the DM features we observe in our Universe \citep{2007Wang, 2012Lovell, 2013Viel, 2014Lovell, 2019Fitts}.
The main approaches groups have taken to constraining WDM particle masses are through comparisons to the observed satellite luminosity function in the Local Group \citep{2014Kennedy, 2017Murgia, 2021Nadler}, comparisons with the LyA forest from spectra of high redshift quasars \citep{2013Viel, 2017Irsic}, and comparisons of the properties of small halos in strongly lensed systems \citep{2019Ritondale, 2019Gilman}.
From the low mass end of the parameter space, combined approaches that utilize multiple methods have produced the strongest constraints on the WDM particle mass.
\cite{2021NadlerA} used a combined analysis of gravitationally lensed systems and satellites around the MW to place the tightest constraint to date of 9.7 keV.
\cite{2021Enzi} performed another combined analysis which also includes the LyA forest and found a lower limit of 6.0 keV.
For a homogeneous population of WDM, a particle mass of $\sim$2 keV is required to match the observational properties of dwarf galaxies \citep{2014Lovell, 2019Fitts}.
This leaves little room for WDM to represent the entire population of DM in our universe.
More recent simulations have employed a mixture of WDM and other DM models to alleviate problems in astrophysical simulations while satisfying constraints on the WDM particle mass \citep{2012Anderhalden, 2016Kamada, 2021Parimbelli}.

Even though WDM works nicely with a machine learning approach to explore its parameters, the constraints placed on the allowed WDM particle mass make it likely ruled out as a viable candidate for the entire population of DM in our Universe \citep{2021Enzi, 2021NadlerA}.
Hence, the DM models we explore here may not be truly physical models that match the properties of the DM we observe in our Universe.
However, exploring the WDM parameter space through a field-level approach with machine learning can provide a more detailed understanding of how varying the DM particle mass affects galaxy formation.
We can then extend this approach to more complex models that may better represent the DM in our universe.

In this paper, we infer the value of DM parameters using information from the field itself. 
We make use of N-body simulations and consider WDM models in this first work and leave the analysis of other DM models and the inclusion of astrophysical processes with hydrodynamic simulations for future work. 
This paper is organized as follows.
In Section \ref{sec:methods} we introduce our simulation suite, WDM models, and CNN architecture.
In Section \ref{sec:results} we present our results from this investigation, focusing on inferring WDM particle masses and cosmological parameters from simulated images.
In Section \ref{sec:discussion} we discuss our results and focus on how the neural network is making its inference to learn what additional information we can extract from astrophysical data.
Finally, in Section \ref{sec:conclusion} we present our conclusions.

\section{Methods}
\label{sec:methods}

In this paper, we employ cosmological N-body simulations with different WDM masses. 
We then use CNNs to perform field-level likelihood-free inference on the value of the WDM mass from 2D images that depict the total matter density field.
In this section, we describe the simulation suites that we employ (\S\ref{sec:SimulationSuite}), the generation of images from the matter density field (\S\ref{sec:ImageCreation}), the architecture of our CNNs (\S\ref{sec:architecture}), the training and validation of our networks (\S\ref{sec:TrainingValidation}), and the testing of our models (\S\ref{sec:testing}).

\subsection{Simulation Suites}
\label{sec:SimulationSuite}

In this paper, we introduce two dark matter only (DMO) simulation suites where we vary the WDM particle mass.
The two simulation suites are distinguished in that one is carried out with fixed cosmological parameters, while the other allows for some variation in the assumed cosmology. 

\begin{table}
    \centering
    \begin{tabular}{l|llll}
        \hline 
        Suite Name  & \# of     & Resolution    & $\Omega_m$ & $\sigma_8$  \\
                    & Sims   &               &            &  \\
        \hline
        Varied Cosmology     & 1000 & 256$^3$ & 0.1-0.5 & 0.6-1.0 \\
        Fixed Cosmology (LR) & 200  & 128$^3$ & 0.302   & 0.839 \\
        Fixed Cosmology (MR) & 200  & 256$^3$ & 0.302   & 0.839 \\
        Fixed Cosmology (HR) & 200  & 512$^3$ & 0.302   & 0.839 \\
        \hline 
    \end{tabular}
    \caption{
    List of simulation parameters for each of our simulation suites.
    \textit{Col 1} gives the name of the suite. LR, MR, and HR refer to low, medium, and high resolutions respectively.
    \textit{Col 2} gives the number of unique simulations in each suite, each with its own random initial density perturbations and WDM particle mass between 2.5 and 30 keV.
    \textit{Col 3} gives the simulation resolution for each simulation with the number of DM particles included in each box.
    \textit{Col 3} and \textit{Col 4} give the cosmological values for each simulation suite where each simulation in the varied cosmology suite has its cosmological values chosen from the ranges given.
    }
    \label{tab:simulations}
\end{table}

Initial conditions are generated at redshift $z=127$ using the \textsc{ngenic} code with the second-order Lagrangian perturbation theory equations~\citep{Springel2019}. We account for the suppression in the initial matter power spectrum due to the WDM streaming velocities through $P_{\rm WDM}(k)=\beta P_{\rm CDM}(k)$,
\begin{align}
\label{eq:wdm}
    \beta &= \left( \left( 1 + (\alpha k)^{2.4} \right)^{-5.0/1.2} \right)^2 \\
    \alpha &= 0.048 \left(\frac{m_{\mathrm{th}}}{1 \mathrm{keV}}\right)^{-1.15} \left( \frac{\Omega - \Omega_b}{0.4} \right)^{0.15} \left(\frac{h}{0.65} \right)^{1.3}  
\end{align}
where $\Omega_m$ is set to our fiducial value of 0.302 and $\Omega_b$ is set to zero because these are DMO simulations.
$m_{\mathrm{th}}$ is the WDM particle mass in keV.
$k$ is the wavenumber measured in h/Mpc where h is 0.6909.
See \cite{2001Bode} for more details.

The initial conditions include random kicks to velocities of the initial DM distribution to simulate the thermal relics that would be present from the WDM.
These kicks ascribe bulk motion to the DM velocities that arise from atomic-level particle motions present from the WDM free streaming velocities.
The magnitude of these kicks is $\sim$6\% for our warmest WDM simulations.
We expect our CNN models to be able to detect this level of variation, so we include these velocities when making our initial conditions.

All simulations have been run with the \textsc{arepo} code \citep{Springel2010, Springel2019, Weinberger2020} that evolves the particles in a periodic cosmological box of $25~h^{-1}{\rm Mpc}$ down to $z=0$. Because we employ DMO simulations, our initial conditions are evolved considering only the influence of gravity.
Gravity is solved for using a TreePM grid ~\citep{Bagla2002} using an assumed gravitational softening of 1.22 ckpc/h and a maximum physical softening of 0.61 kpc/h for our fiducial model.
Snapshots are generated and saved at redshifts $z=\{5,4,3,2,1,0\}$.

\subsubsection{Varied cosmology suite}

\begin{figure}
 	\includegraphics[width=\columnwidth]{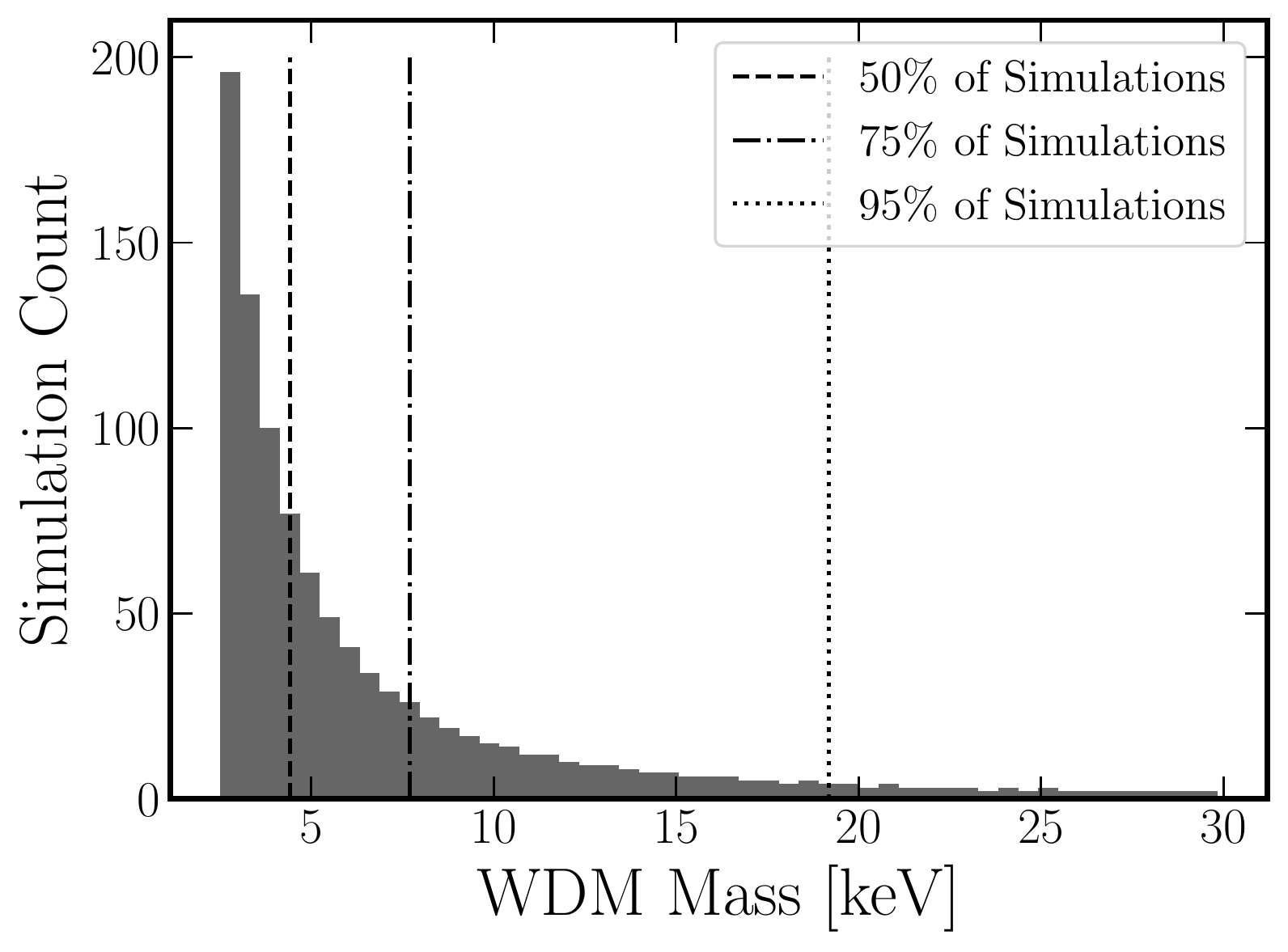}
    \caption{Histogram of 1000 unique WDM particle masses used in the varied cosmology simulation suite. The particle masses were sampled in tandem with $\Omega_m$ and $\sigma_8$ when creating a latin-hypercube. We sample the particle masses from an inverse distribution to increase the proportion of the training set with a strong signal for the model to learn. The fixed cosmology suite shares the same distribution of particle masses, but only contains 200 simulations.}
    \label{fig:wdm_hist}
\end{figure}

The varied cosmology simulation suite has 1,000 unique simulations, where the value of $\Omega_m$, $\sigma_8$, and the WDM particle mass is different in each simulation and the value of these parameters are organized in a latin-hypercube.
The initial random seed is also different for each simulation.
This parallels the approach used in the original CAMELS project \citep[]{CAMELS}. 
The cosmological parameters are sampled uniformly in the ranges $0.1 < \Omega_m < 0.5$ and $0.6 < \sigma_8 < 1.0$.
These ranges were chosen to match the CAMELS simulations which sampled a large section of parameter space to mitigate prior effects on the neural network results.
The WDM particle mass is sampled uniformly from an inverse distribution of particle masses (i.e. uniform in $1/m_{\mathrm{WDM}}$) with $m_{\mathrm{WDM}}$ ranging from 2.5 keV to 30 keV (see Figure \ref{fig:wdm_hist}).
The small number of simulations at high WDM particle masses may limit the model's effectiveness for these masses.
Since the CNNs make predictions in inverse keV space where the particle masses are uniformly sampled, this distribution should not bias the training process toward a specific mass.

Our fiducial resolution simulations contain 256$^3$ particles which provide a mass resolution of $7.81 \times 10^7 (\Omega_m/0.302) h^{-1} M_\odot$.
At this resolution, we employ a maximum physical gravitational softening of 1.22 kpc/h.

\subsubsection{Fixed cosmology suite}

The fixed-cosmology suite shares most characteristics with the varied cosmology suite, except that it contains 200 simulations with $\Omega_m$ fixed at 0.302 and $\sigma_8$ fixed at 0.839, consistent with Plank data \citep{2016Planck}. 
As for the varied cosmology simulation suite, each simulation has a different value of the initial random seed and thus includes a different initial realization.
Additionally, for the fixed cosmology suite, each simulation is paired with two additional simulations that have the same cosmology, initial random seed, and WDM particle mass but have been run at lower resolutions.
The high/medium/low resolution simulations contain $512^3/256^3/128^3$ particles with masses $9.74 \times 10^6 h^{-1} M_\odot$ / $7.81 \times 10^7 h^{-1} M_\odot$ / $6.24 \times 10^8 h^{-1} M_\odot$.
This choice allows for an even-handed comparison between the resulting CNNs that are trained on the simulation suites of varied resolution.

\subsection{Image creation}
\label{sec:ImageCreation}

Projected images of the matter density field (i.e. images of the matter surface density) are created by taking slices of the three-dimensional matter density field and then projecting them along one axis. The three-dimensional matter density field is first calculated as follows. First, we compute the distance of each dark matter particle to its closest 32$^{nd}$ neighbor, $R_{32}$.
Next, each dark matter particle is considered to be a sphere with radius $R_{32}$ and uniform density. Finally, the mass in the spheres is assigned to a 3D regular grid using the \textsc{voxelize} code\footnote{\url{https://github.com/leanderthiele/voxelize}}.
We also trained a model using a nearest grid point assignment scheme and found similar results to our method, suggesting that our results are not strongly sensitive to the chosen image creation method.

For a $N\times N$ pixel image, we calculate the three-dimensional matter densities on a grid with $N\times N\times N$ voxels over the whole simulation volume. We then take a slice with dimensions $N\times N \times d$, where $d$ is chosen to represent a length of $5~h^{-1}{\rm Mpc}$. Finally, the mass in every voxel is projected along the smallest axis and the image is created. Note that we compute mass surface density, so we divide the mass of every pixel by $L^2$, where $L$ is the simulation box size. We take five non-overlapping slides per axis. Thus, we have 15 images per simulation.

During the training process, we also produce additional images which are rotated 90$^{\circ}$, 180$^{\circ}$, and 270$^{\circ}$ from the original and an additional set of images that are the same rotations of an image reflected over the y-axis.
This produces an additional 7 images for each of the 15 unique images discussed above.
For the varied cosmology suite, we have 120,000 images per epoch to train, validate, and test our CNN.
For the fixed-cosmology suite, we instead have 24,000 images. See Figure \ref{fig:images} for example images from our fixed-cosmology simulation suites.

Before the images are given to the CNN, they are first logged and normalized so that all pixel values fall between zero and one.
Normalizing these images is standard practice for training CNNs as it reduces the model's intrinsic bias toward data with larger absolute values.

\subsection{Architecture}
\label{sec:architecture}

We adopt the same CNN architecture as presented in \cite{CAMELS} and used in \cite{2021VillaescusaA, 2021Villaescusa} to train on the CAMELS Multifield Dataset \citep{CMD}.
See Section 3.1.1 of \cite{CMD} for an outline and details of the architecture.
We summarize the key points below.

The CNN is comprised of six convolutional blocks, one additional convolutional layer, and one fully connected block.
Each convolutional block has three convolutional layers, each of which is followed by a batch normalization, and passed through a leaky ReLU nonlinear function.
The first and second convolutional layer in each block has a 3x3 filter size with a step of one.
The last convolutional layer has a 2x2 filter size and a step of 2 to reduce the size of the image as it is passed through the layers.
By the final convolutional block, the resulting image is reduced to a single pixel with multiple layers (the number of which depend on the hyperparameters for the particular model).

In the fully connected block, the resulting tensor from the above convolutional blocks is flattened, a certain percentage of neurons are then dropped (depending on the dropout rate hyperparameter), and then passed through a fully connected layer.
The results are then passed through another leaky ReLU nonlinearity, drop out, and fully connected layer to produce two values, the mean and standard deviation of the predicted posterior distribution.

The architecture is adapted slightly for images with different pixel resolutions.
For images with 512$^2$ pixels, an additional convolutional block is added at the beginning of the architecture described in \cite{CMD}.
Similarly, for an image with 128$^2$ pixels, the first convolutional block is removed.
In this way, the image size is reduced to a single pixel once it passes through the entire CNN.
We do not find that the additional convolutional blocks present for higher resolution images significantly affect any results we present here.

Our model takes as input a 2D image and outputs two numbers, the posterior mean ($\mu_i$) and standard deviation ($\sigma_i$) of the WDM particle mass. To achieve that, we made use of this loss function 

\begin{equation}
\label{eq:loss}
    \mathcal{L} = \left( \sum_{i \epsilon \mathrm{batch}} (\theta_i - \mu_i)^2 \right) + \left( \sum_{i \epsilon \mathrm{batch}} ((\theta_i - \mu_i)^2 - \sigma_i^2)^2 \right)
\end{equation}

where $\theta_i$ is the WDM particle inverse mass of the map $i$. We refer the reader to \cite{moments_network, CMD} for a full description of this loss function.

\subsection{Training and Validation}
\label{sec:TrainingValidation}

To train and validate our model, we first split our image sets into 90\% training, 5\% validation, and 5\% testing.
For our varied cosmology simulation, this is 900 simulations for the testing set (108,000 images), and 50 simulations for each of the validation and testing sets (6,000 images each).
For our fixed cosmology suite, this is 180 simulations for the testing set (21,600 images, and 10 for each of the validation and testing sets (1,200 images each).
For the fixed-cosmology suite, we include 20 additional simulations and their corresponding images in the testing set that were not included in the training or validation sets.
These additional simulations allow for more populated result plots where trends are not otherwise apparent.

We tune the value of the hyperparameters, the learning rate, the weight decay, the dropout rate, and the number of hidden channels in the CNNs, using \textsc{optuna} \citep{optuna} for 50 trials. See Table \ref{tab:hyperparameters} for details on parameter ranges and optimal values.
During the tuning process, the model with the best validation loss is selected and saved for testing. Each trial trains the model for 200 epochs where in each epoch the model sees each image in the training set.

\begin{table}
    \centering
    \begin{tabular}{l|llll}
        \hline
        Name & Min & Max & Fiducial Fixed Value & Fiducial 
        Varied Value  \\
        \hline 
        LR & 1e-5 & 5e-3 & 3.4e-3 & 4.8e-3 \\
        WD & 1e-8 & 1e-1 & 5.7e-3 & 2.3e-7 \\
        DR & 0.0  & 0.9  & 0.68   & 0.60 \\
        H  & 6    & 12   & 11     & 11 \\ 
        \hline 
    \end{tabular}
    \caption{List of hyperparameters for our CNNs (col 1), the minimum (col 2) and maximum (col 3) value over which we vary them during tuning, and the chosen values for our fiducial fixed (col 4) and varied (col 5) models.
    The hyperparameters tuned for these models are learning rate (LR), weight decay (WD), decay rate (DR), and the number of hidden layers (H).
    See Section \ref{sec:TrainingValidation} for more details.
    We tune the hyperparameters with \textsc{optuna} over 50 trials where all values are varied simultaneously.
    The chosen values in columns 3 and 4 were chosen from the model with the lowest validation loss.
    }
    \label{tab:hyperparameters}
\end{table}

\subsection{Testing}
\label{sec:testing}

To test our models, we first compare the loss value for the images from our testing set to the images in our validation set to ensure that our image sample is not significantly biased in a way that will affect our results.
The posterior mean and standard deviation are output by the CNN (see the previous section for details) and compared to the true WDM particle mass.
Results are produced in inverse keV space as that limits the span of reasonable values of the parameter space from 0 (CDM) to 1 (WDM) instead of 1 (WDM) to infinity (CDM).
To present our results more intuitively, we transform the mean and errors from inverse (keV$^{-1}$) to non-inverse (keV) units.
In practice, this does not change the relative difference between the true and predicted values, but it does compress and extend some of the data in an absolute sense.

To present our results, we display the posterior mean vs the true WDM particle value (both in keV) of a single image taken from each simulation in our test set.
Points that fall on a one-to-one line are accurate predictions from the CNN.
With this method, it can be ambiguous as to where the transition is from accurate predictions (for warmer models) to inaccurate predictions (for colder models).

To mitigate this ambiguity, we provide an analytical technique to quantify where this transition happens and to what WDM particle mass a given model can make accurate predictions with a certain confidence level.
In this method, we begin by taking a first pass over the data with an MCMC approach to fit a two-parameter model to our data.
The two parameters are the point at which a one-to-one line transitions to a flat line and the height of the flat line.
The transition parameter marks where our model can no longer make accurate predictions and instead guesses an average value.
The second parameter is not used directly in our analysis but is included to allow the MCMC model more flexibility in choosing the transition point.
To obtain a more robust measurement, we also include data points from all unique images in the testing set (not including rotations or reflections) during the MCMC fit.
These additional data points are not included in any figures shown in Section \ref{sec:results}.

Since some data points that are classified as accurate predictions may be indistinguishable from the distribution of inaccurate predictions, we define a new metric to measure where our model can make accurate predictions that are statistically separate from an inaccurate prediction.
To do this, we begin by separating the data points used in the MCMC fit into two categories based on if they fall on the one-to-one line (accurate predictions) or not (inaccurate predictions).
We then take the distribution of inaccurate predictions and calculate the standard deviation around that average.
We define the upper limit of accurate predictions our model can make to be the simulation where the predictions for all unique images in that simulation are at least 2$\sigma$ away from the average inaccurate predictions.
The error bar on this upper limit is taken to be the distance between the best-measured upper limit and the next true WDM mass.
This method is used to compare the resolution- and redshift-dependent models in Section \ref{sec:evaluation}.

\section{Results}
\label{sec:results}

In this section, we present the results of our investigation.
We begin by presenting the images and results of our fiducial setup and then investigate how modifications to this setup affect our results.
Our fiducial setup consists of our fixed cosmology simulation suite at 512$^3$ particle resolution and 512$^2$ pixel resolution in our images.
The various modifications we will explore are varied simulation resolution, varied image resolution, varied redshift, and varied cosmology.

\begin{figure*}
        \includegraphics[width=.8\textwidth]{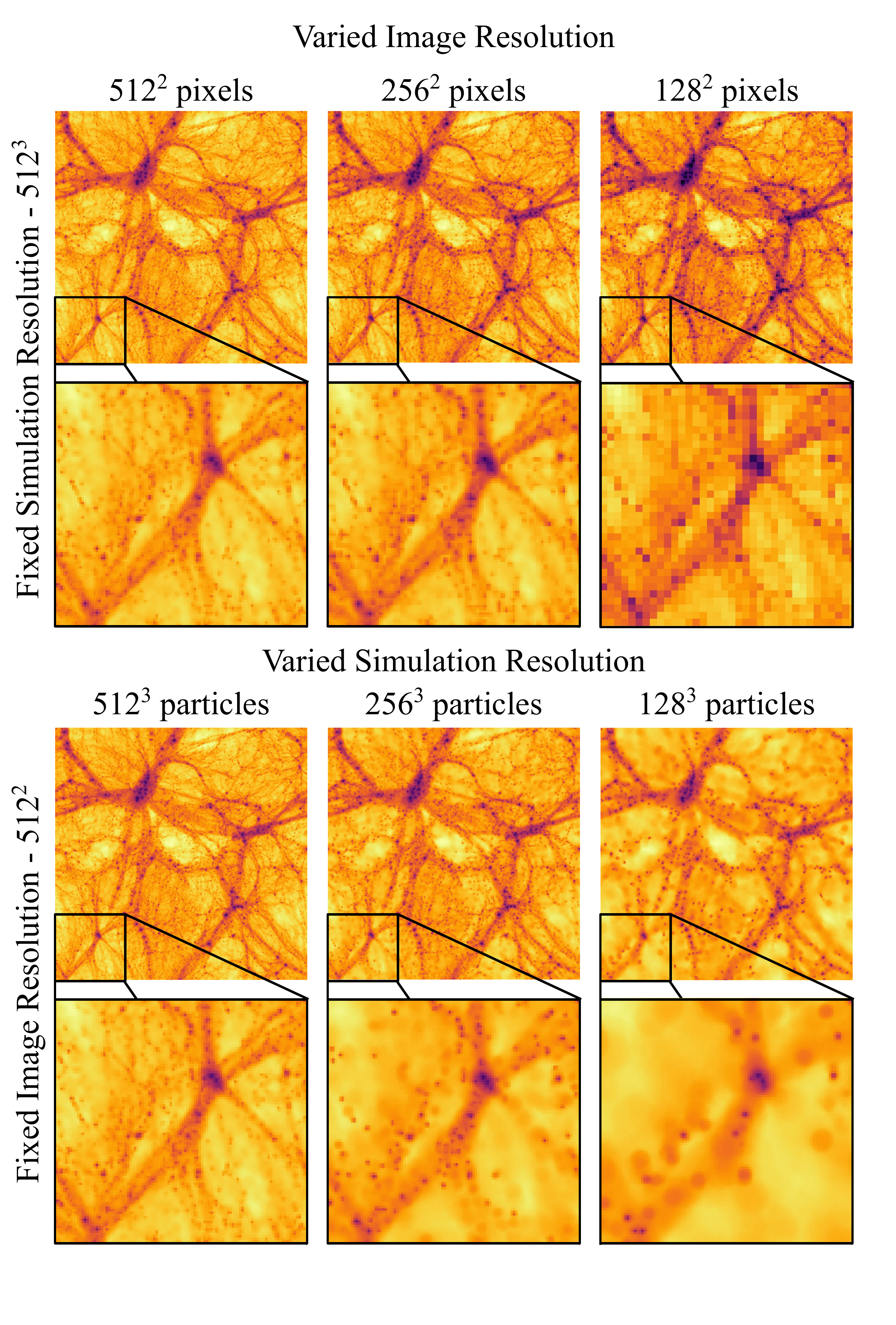}
    \caption{A comparison of images created at different image resolutions (top) and simulation resolutions (bottom). 
    In the top section of this plot, we compare a single image created from the same high-resolution simulation, but at different image resolutions: 512$^2$ (left), 256$^2$ (middle), and 128$^2$ (right) pixels. 
    In the bottom section of the plot, we compare a single image created from simulations with different mass resolutions but presented at the same image resolution.
    Each image is shown at a fixed dynamic range for comparison.
    The simulations contained 512$^3$, 256$^3$, and 128$^3$ particles with a mass resolution of $9.74 \times 10^6$, $7.81 \times 10^7$, $6.24 \times 10^8$  $h^{-1} M_\odot$ respectively.
    For both the simulation and image resolution variations, high-density areas remain relatively unchanged while low-density areas and small structures vary significantly across resolutions.}
    \label{fig:images}
\end{figure*}

Figure \ref{fig:images} shows a sample of images used for training our models.
The top section shows image dependence on the number of pixels.
From left to right, the image resolutions are 128$^2$, 256$^2$, and 512$^2$ pixels.
All images are from the same simulation with 512$^3$ particle resolution.
In the second row, we show an enlarged section of the simulation field.
The particular image and enlarged region we display were selected to be representative of most images in the dataset but were not chosen for any particular features.
With decreased resolution, low-density areas with smaller structures are the most affected.

The bottom section of Figure \ref{fig:images} shows a sample of images taken from different simulation resolutions at a fixed image resolution of 512$^2$ pixels.
From left to right, the simulation resolutions are 128$^3$, 256$^3$, and 512$^3$ particles with a mass resolution of $6.24 \times 10^8$, $7.81 \times 10^7$, $9.74 \times 10^6$ $h^{-1} M_\odot$ respectively.
We also show the same enlarged region as above for each image in the bottom row.
Similar to the image resolution, the largest changes are most apparent in the low-density areas.

\subsection{WDM Mass Inference}
\label{sec:fixedCosmo}

\begin{figure}
	\includegraphics[width=\columnwidth]{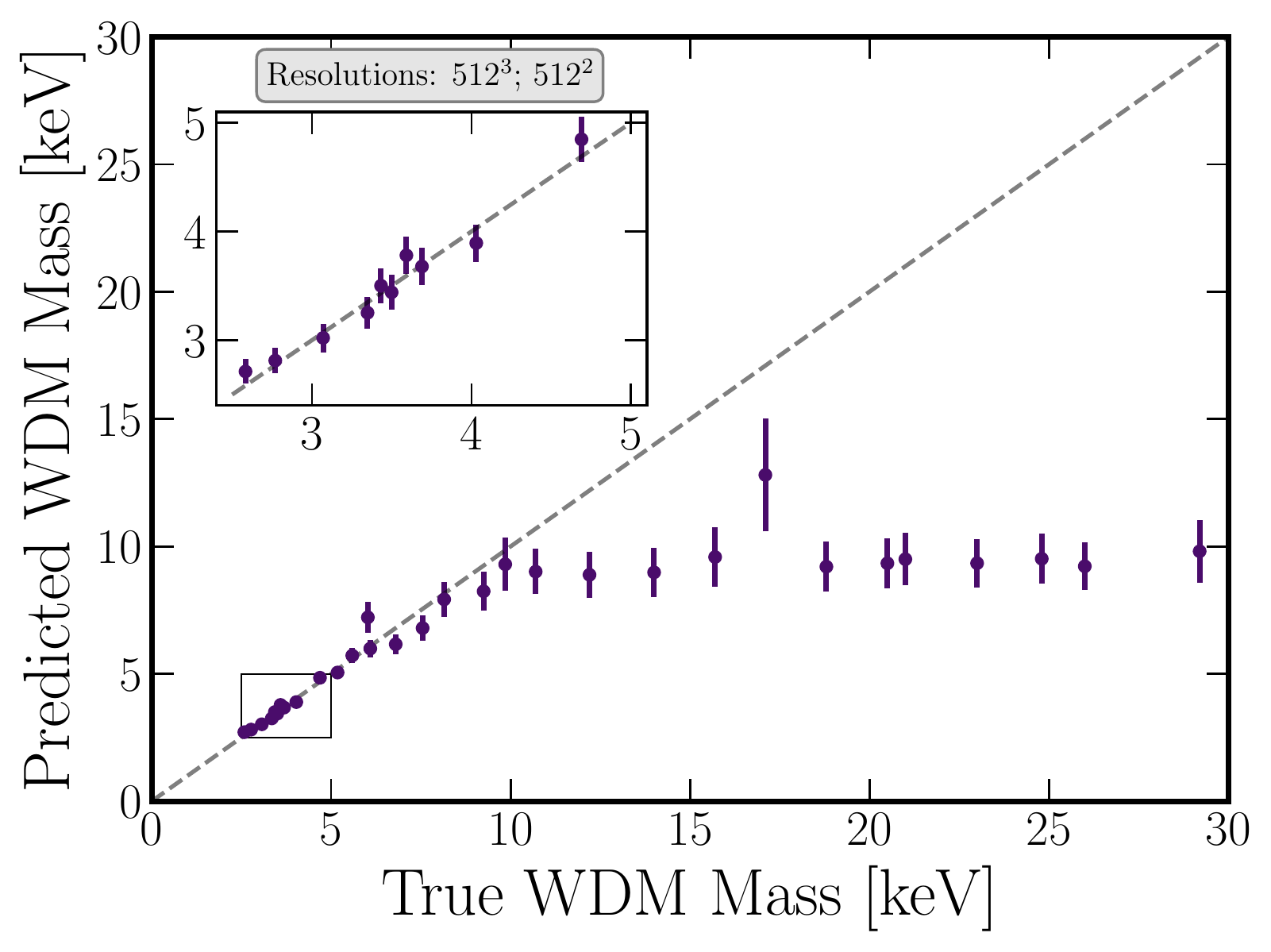}
    \caption{Results from our CNN model trained on our fiducial fixed cosmology simulation suite at a simulation resolution of 512$^3$ particles ($9.74 \times 10^6$ $h^{-1} M_\odot$) 
    and 512$^2$ pixel image resolution. 
    The x-axis shows the WDM particle mass in keV that was used to create each simulation.
    The y-axis shows the WDM particle mass that is predicted by the CNN from a single image taken from each simulation in our test set.
    A one-to-one line (dashed) is shown where predictions match the true value.
    Predictions match the true value well up to 10 keV,
    with an uncertainty of $\pm$1 keV,
    where the model can no longer distinguish between different WDM models and guesses a constant value of $\sim$10 keV.}
    \label{fig:results_fixed}
\end{figure}

Figure \ref{fig:results_fixed} shows the results of our fiducial model (512$^3$ particles; 512$^2$ pixels).
Each data point shows the prediction from the trained CNN on a single image from each simulation in our test set.
The x-axis shows the true WDM particle mass in keV that was used to create the simulations and images.
The y-axis shows the WDM particle mass predicted by the CNN after it has completed the entire training and hyperparameter tuning process.
A one-to-one line (dashed) is shown where the predicted value matches the true value.
We include an enlarged view of the 2.5 to 5.0 keV WDM mass range in the upper left section of each plot to better view the points with small uncertainties.

For our fiducial model, the predictions agree very well with the true WDM particle mass up to 10 keV.
After 10 keV, the model predicts a constant value of $\sim$10 keV up to the end of our sample at 30 keV.
There are two main outliers presented here: the prediction that lies above the one-to-one line near 6 keV and the prediction that lies above the guessed values near 17 keV.
These outliers can likely be explained by anomalies in the single image that was used to make the predictions, where a different image from the same simulation would likely give a more expected result.

Outliers and the systematic deviation from the one-to-one line at high particle masses are discussed more in Section \ref{sec:outliers}.
Additionally, many points near 10 keV may be accurate in that the predicted value is near the true particle mass, but if taken out of context, may be indistinguishable from a 30 keV image which also has a predicted value of 10 keV.
We will discuss these points and how to faithfully evaluate these models in Section \ref{sec:evaluation}.

\subsection{Model Variations}
\label{sec:variations}

In this section, we discuss how variations to our fiducial model, through image resolution, simulation resolution, redshift, and cosmology impact their predictive power.
We start by providing a selection of representative models to compare to our fiducial model presented above.
We then use the method outlined in Section \ref{sec:testing}, to quantitatively evaluate and compare these models to uncover how these variations affect their predictive power.

\begin{figure}
 	\includegraphics[width=\columnwidth]{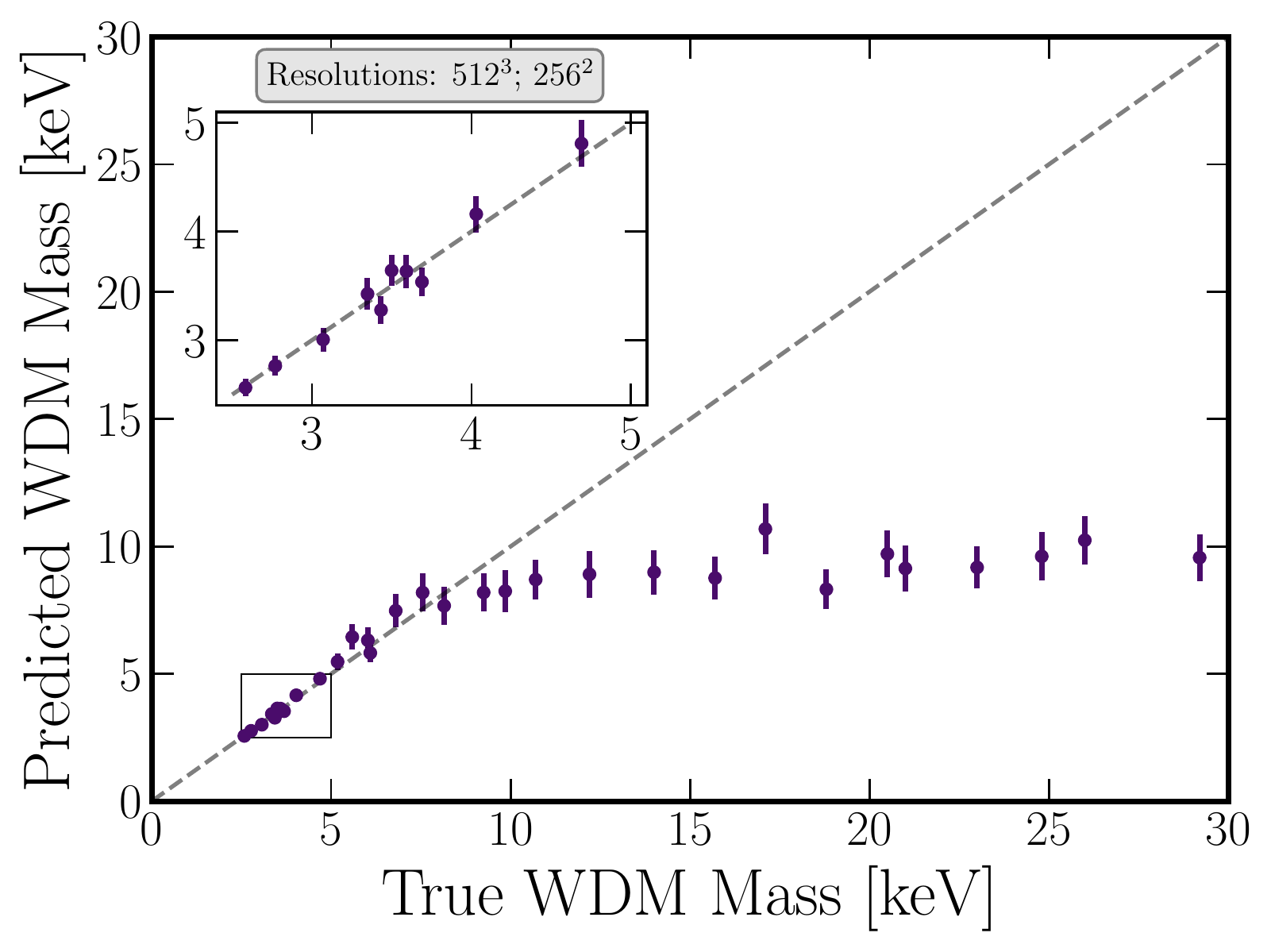}
  	\includegraphics[width=\columnwidth]{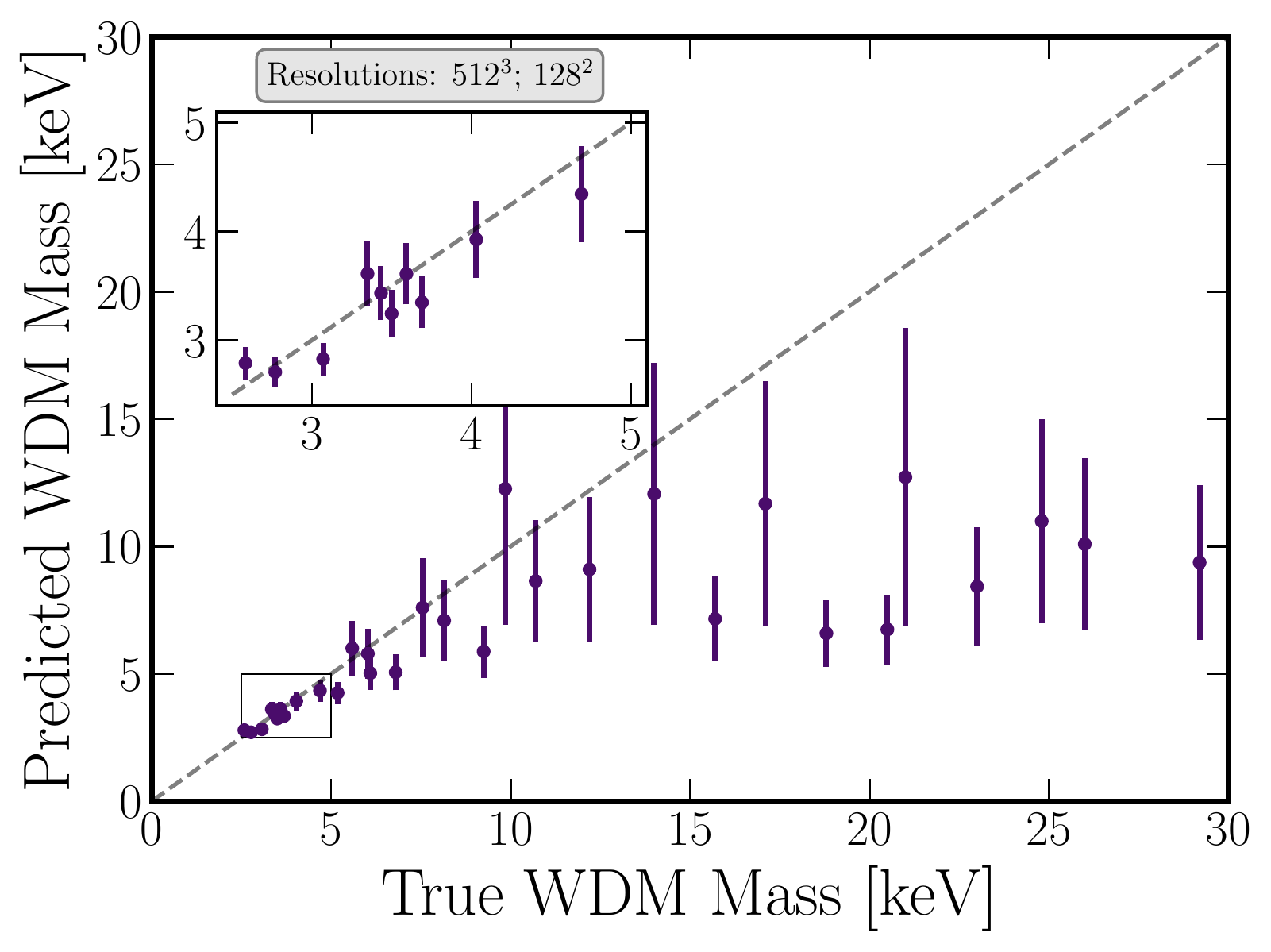}
    \caption{Results for models trained on images from the fixed-cosmology suite with 128$^2$ (bottom) and 256$^2$ (top) image resolution. 
    The simulation resolution is fixed at our fiducial model of 512$^3$ particles, or a mass resolution of $9.74 \times 10^6$ $h^{-1} M_\odot$.
    For images with 256$^2$ pixel resolution (top), the model can make accurate predictions up to 8 keV with an average uncertainty and error of 0.3 keV and 0.2 keV respectively.
    At 128$^2$ pixel resolution (bottom), the model can make accurate predictions up to 7 keV with an average uncertainty and error of 0.4 keV and 0.4 keV respectively.
    Decreasing the image resolution decreases the WDM particle mass the model can accurately distinguish while increasing the uncertainty and error on those predictions. 
    We note that while some data may be considered accurate, it can be indistinguishable from a higher particle mass model.
    A more robust way to evaluate the maximum mass to which our models can accurately predict WDM models is presented in Section \ref{sec:evaluation}.
    }
    \label{fig:im_res}
\end{figure}

In Figure \ref{fig:im_res}, we present results from three different models with varied image resolution and a fixed simulation resolution (512$^3$ particles).
Each panel represents a distinct model that has been trained on a unique set of images for the given resolution, but each of which was created from the same simulation suite.
The image resolutions are 256$^2$ (top) and 128$^2$ (bottom) pixels.
Figure \ref{fig:results_fixed} shows the results from the 512$^2$ pixel model and the top section of Figure \ref{fig:images} for an example of images that were used to train each of these models.

The top panel shows a similar plot to the fiducial model shown in Figure \ref{fig:results_fixed}, but for images at 256$^2$ pixel resolution.
The fiducial model does well up to 10 keV, with an average uncertainty of 0.3 keV and an average error of 0.3 keV.
This 256$^2$ pixel model can make accurate predictions up to 8 keV.
Up to 8 keV, the model has an average uncertainty of 0.3 keV and an error of 0.2 keV.
This provides a similar uncertainty to our fiducial model and slightly better error but does not extend to as high of particle masses.

The bottom panel shows results from our model trained with images at 128$^2$ pixel resolution.
This model can make accurate predictions up to 7 keV with an average uncertainty of 0.4 keV and an average error of 0.4 keV.
This represents a doubling in error and uncertainty compared to our fiducial model over the same mass range.
As we noted for our fiducial model, some of these points may be considered accurate, but are indistinguishable from a 30keV image where the model predicts a similar value.
We will discuss these points in Section \ref{sec:evaluation}.

\begin{figure}
 	\includegraphics[width=\columnwidth]{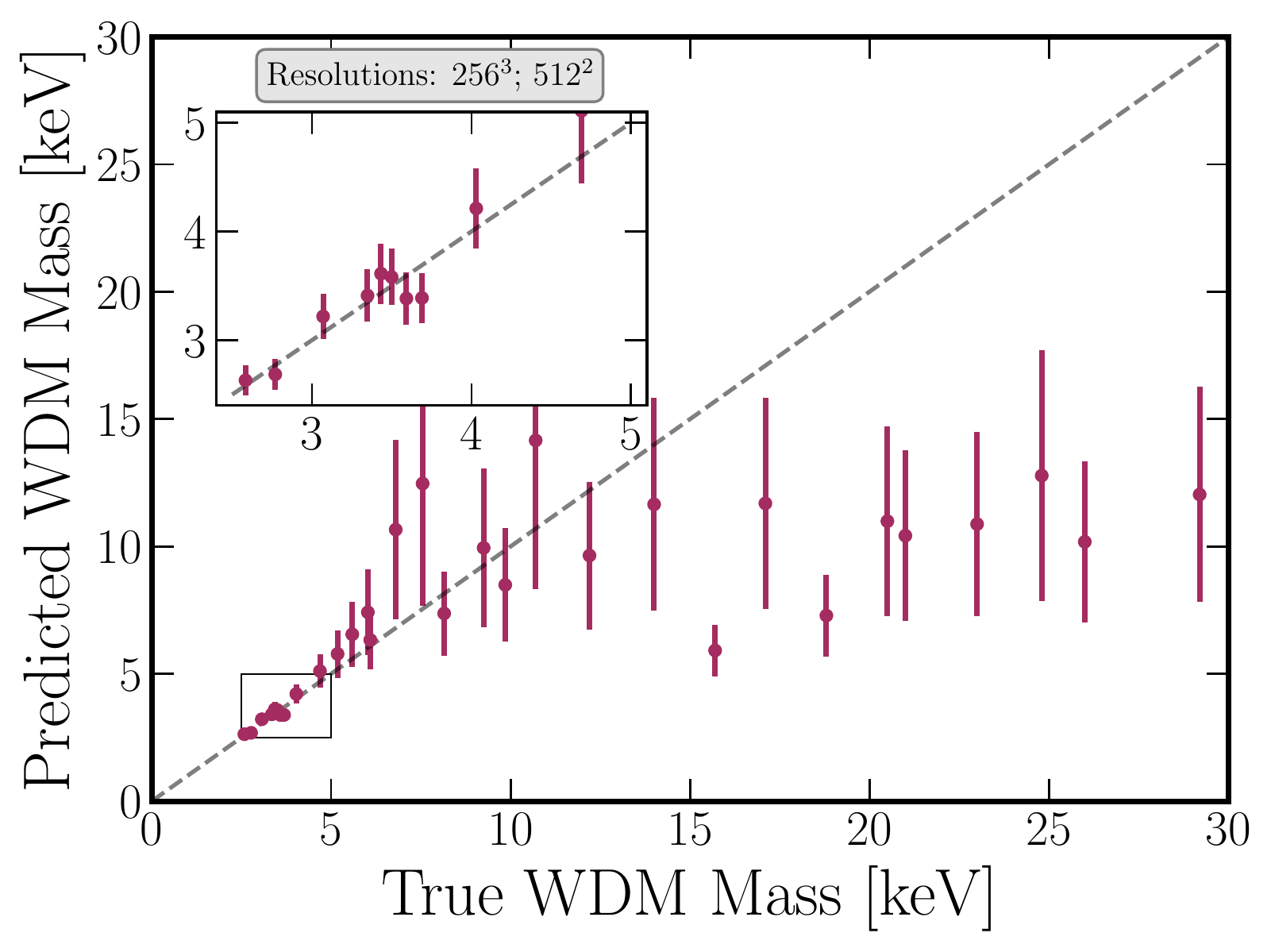}
  	\includegraphics[width=\columnwidth]{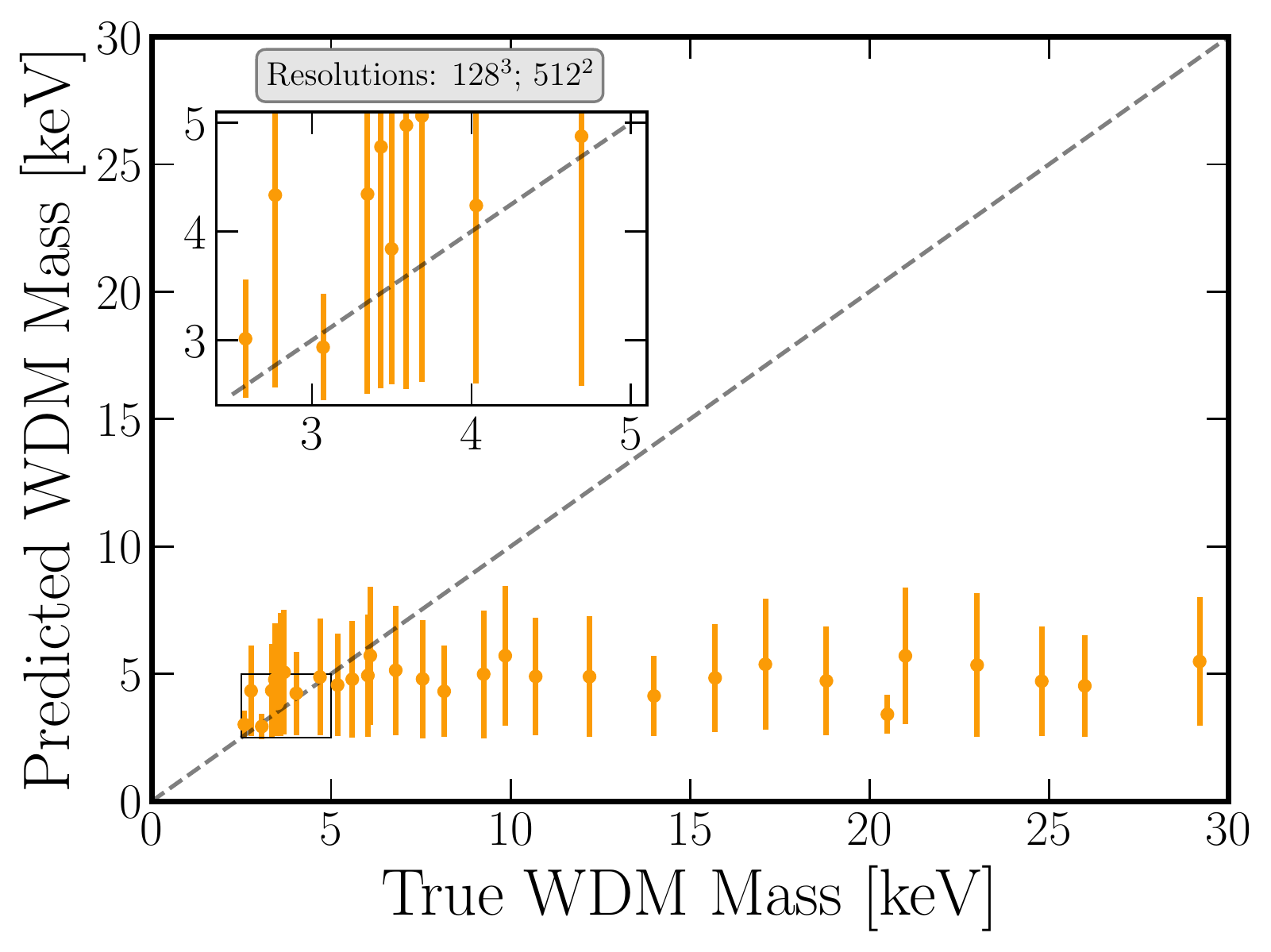}
    \caption{A comparison of results from our 256$^3$ (top) and 128$^3$ (bottom) simulation resolution results. All images used to train and test these models have 512$^2$ pixels.
    The top panel is our fiducial model and the results are the same as Figure \ref{fig:results_fixed}.
    The 256$^3$ simulation resolution model (top) can accurately predict to 7 keV with an average uncertainty and error of 0.8 keV and 0.6 keV respectively.
    The 128$^3$ simulation resolution model (bottom) can accurately predict to 3 keV with an average uncertainty and error of 1.2 keV and 1.0 keV respectively.
    Overall, the predictive power and precision of these models decrease with decreased simulation resolution faster than comparable decreases to the image resolution presented in Figure \ref{fig:im_res}.
    }
    \label{fig:sim_res}
\end{figure}

Figure \ref{fig:sim_res} shows how varying the simulation resolution at a fixed image resolution affects our results.
All images that were used to train and test these models were fixed at our fiducial resolution of 512$^2$ pixels.
The simulation resolutions are 256$^3$ (top) and 128$^3$ (bottom) particles.
These resolutions represent mass resolutions of $7.81 \times 10^7$ and $6.24 \times 10^8$ $h^{-1} M_\odot$ respectively.

The top panel shows the results of our model trained and tested on images taken from a simulation with 256$^3$ particles.
This model can accurately make predictions up to 7 keV with an average uncertainty of 0.8 keV and an average error of 0.6 keV.
This represents an increase in both uncertainty and error of 330\% and 280\% respectively from our fiducial model over the same mass range (see Figure \ref{fig:results_fixed}).
Compared to the top panel in Figure \ref{fig:im_res}, this represents an increase in uncertainty and error of 250\% and 360\% respectively compared to the corresponding model where we decrease image resolution by a factor of 2 from our fiducial setup.

The bottom panel shows the results of our model trained and tested on images taken from a simulation with 128$^3$ particles.
For this model, results are only accurate up to 3 keV, with an average uncertainty of 1.2 keV and an average error of 1.0 keV.
This represents an increase in both uncertainty and error of 1000\% and 1200\% respectively from our fiducial model over the same mass range.
Compared to the right panel in Figure \ref{fig:im_res} where the image resolution was reduced by a factor of 4, this model has an increase in both average uncertainty and error by 800\% and 740\%.

\subsection{Model Evaluation}
\label{sec:evaluation}

Here we will employ the method we outline in Section \ref{sec:testing} to evaluate the maximum WDM particle mass up to which our models can accurately and confidently make predictions.
Specifically, we compare the predictions from a given simulation against a distribution of predictions we know to be incorrect.
The degree (or lack) of separation between these two distributions provides the confidence level upon which we can trust that the predictions are accurate.
We define the predictive power of our models as the highest WDM particle mass where the predicted WDM particle masses are accurate and distinct from our distribution of inaccurate predictions at the 95\% confidence level.
This method of determining predictive power provides a stricter measurement of accuracy than solely relying on uncertainties and errors, but it more faithfully represents which predictions can be trusted.

We use the predictive power from this method as a proxy for the amount of information contained in the images that can be used to distinguish between WDM models.
This then allows us to systematically vary aspects of the data provided to the CNN, such as image resolution, simulation resolution, or redshift, to determine where and when information is present that can distinguish between various WDM models.
See Section \ref{sec:saliency} for a discussion on these variations.

\subsubsection{Resolution}

\begin{figure}
	\includegraphics[width=\columnwidth]{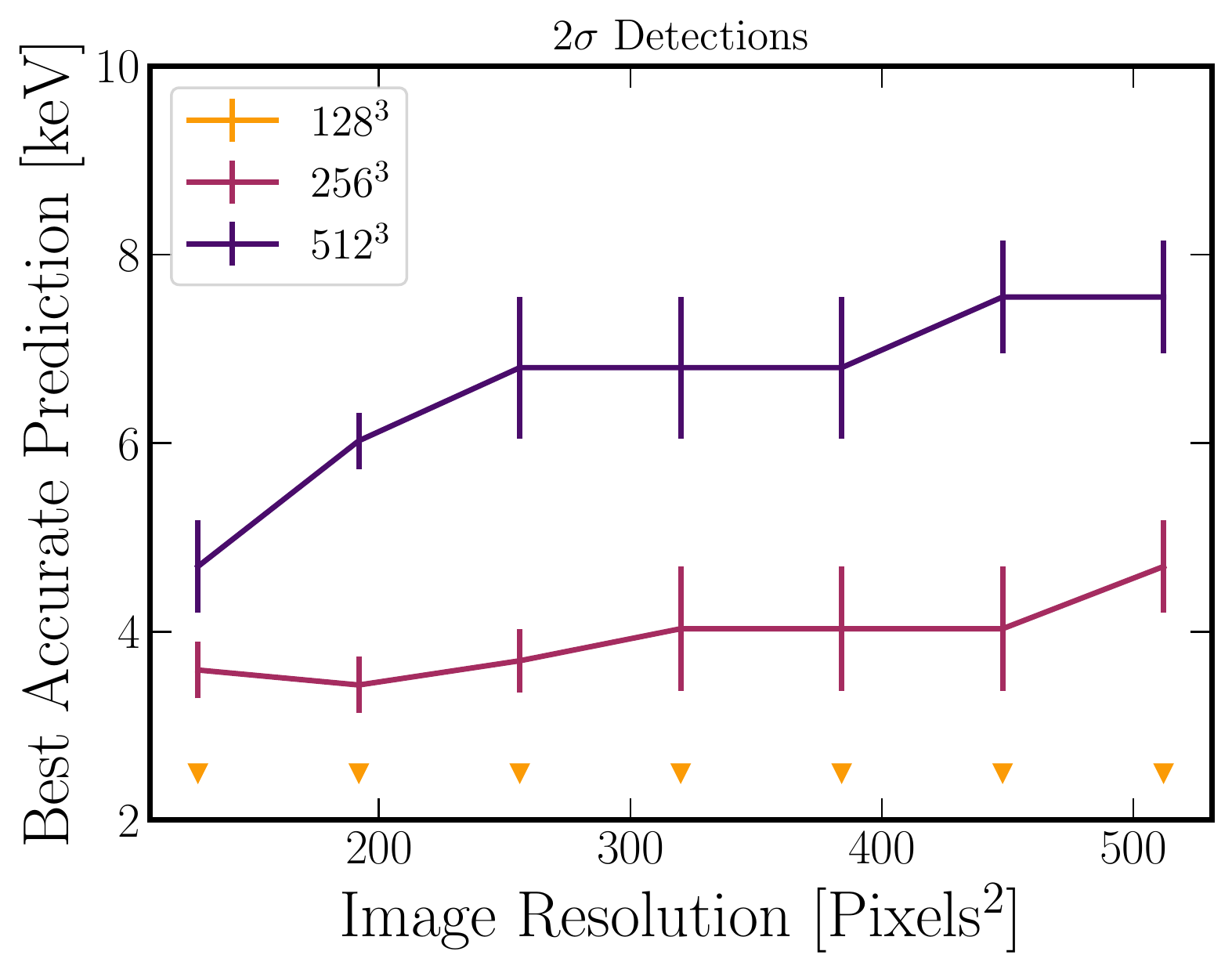}
    \caption{A comparison of results from models trained on different image resolutions (x-axis) and simulation resolutions (lines). 
    We find that at our lowest simulation resolution, the model cannot make any accurate predictions regardless of image resolution. 
    For our medium and high-resolution simulations, we find that the models do continually better as we increase image resolution. 
    For our high-resolution simulations, we find that our models have a large increase in predictive power from 128$^2$ to 256$^2$ pixels.}
    \label{fig:resolution}
\end{figure}

In Figure \ref{fig:resolution}, we compare models trained at different simulation and image resolutions to each other following the method outlined in Section \ref{sec:testing}.
For both Figures~\ref{fig:resolution} and ~\ref{fig:redshift}, we present the maximum WDM mass where the model reaches a 95\% confidence level that its predictions are distinct from a random guess.

In this figure, we present the results from three different simulation resolutions: 128$^3$, 256$^3$, and 512$^3$ particles.
For each of these simulation resolutions, we present results from seven different image resolutions ranging from 128$^2$ pixels up to 512$^2$ pixels, with additional resolutions at every additional 64 pixels per side in between.
For models that are trained on images that are not a factor of two (ex. 192$^2$ pixels), we pad the images with zeros until they contain a number of pixels that is a factor of two along each axis.
This extra padding allows us to use our models on images at intermediate resolutions but is not expected to affect our results.

We find that for our low-resolution simulation, no model can make any predictions to the required confidence level for the image resolutions and WDM particle masses we explore here.
For our medium-resolution simulation, we find an average predictive power of $\sim$4 keV with a marginal increase in predictive power with image resolution.
For our high simulation resolution models, we find a large increase (from 4.7 to 6.8 keV) in predictive power when going from 128$^2$ to 256$^2$ pixels.
From 256$^2$ to 512$^2$ pixels, we find a marginal increase in our predictive power.

Overall, we find that simulation resolution provides a much larger increase to the predictive power than changing image resolution.
The large increase in predictive power for our high-simulation resolution model from 128$^2$ to 256$^2$ pixels suggests a size scale that is necessary to resolve at this resolution to make accurate predictions.
We will discuss this more in Section \ref{sec:saliency}.
We also note that there is not an uncharacteristic and systematic jump in predictive power from 128$^2$ to 192$^2$, 192$^2$ to 256$^2$, or 448$^2$ to 512$^2$ pixels suggesting that the additional layer included in the architecture to account for the larger images does not significantly affect our results.

\subsubsection{Redshift}
\label{sec:redshift}

\begin{figure}
	\includegraphics[width=\columnwidth]{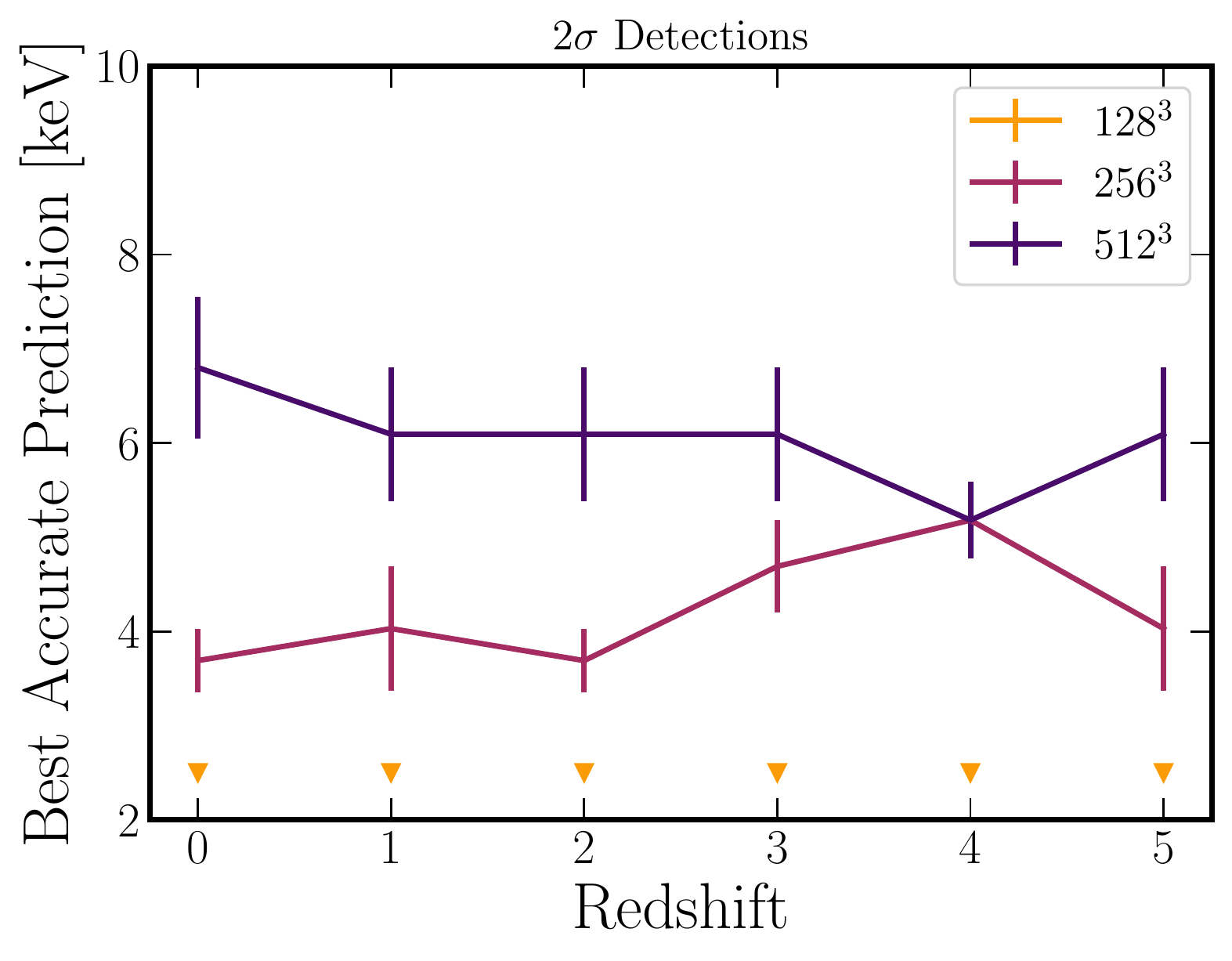}
    \caption{A comparison of results from models trained on images from z=0 to z=5 in the fixed-cosmology simulation suite for each of our three simulation resolutions. 
    Each model is trained with images at a resolution of 256$^2$ pixels and evaluated according to the method outlined in Section \ref{sec:testing}.
    For the low-resolution models, we find that the models cannot make any predictions with a 95\% cl. at any redshift, up to z=5.
    For the medium resolution models, we find no increase in predictive power up to z=2, after which the predictive power increases from 4.0 keV to 5.2 keV by z=4, then decreases again to 4.0 at z=5.
    For our high-resolution models, we find little evolution in the mass to which our model can accurately make predictions with redshift.
    There is a slight increase in the predictive power at z=0 and a slight decrease at z=4.
    Overall, we find no systematic dependence on our models' predictive power across redshift up to z=5.
    }
    \label{fig:redshift}
\end{figure}

\begin{figure*}
	\includegraphics[width=.33\textwidth]{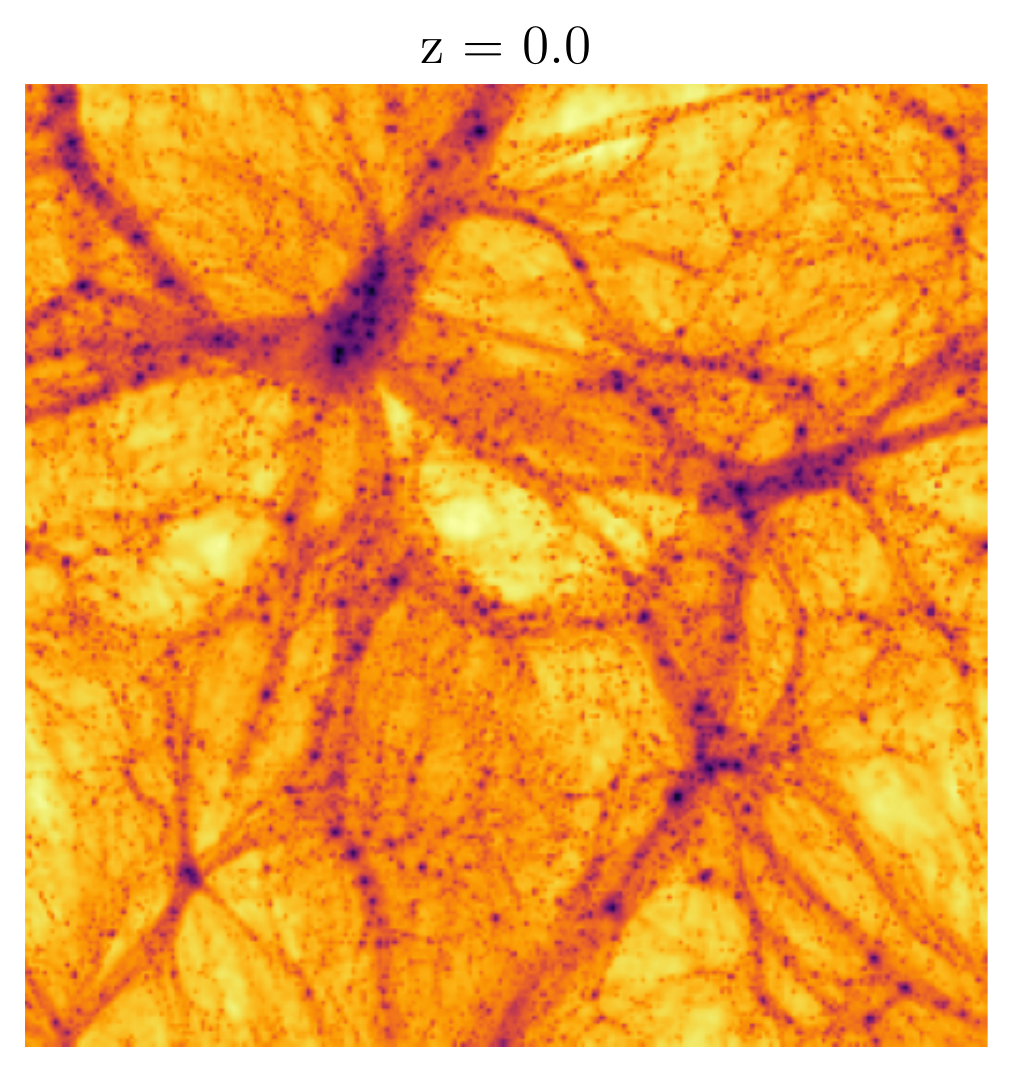}
       \includegraphics[width=.33\textwidth]{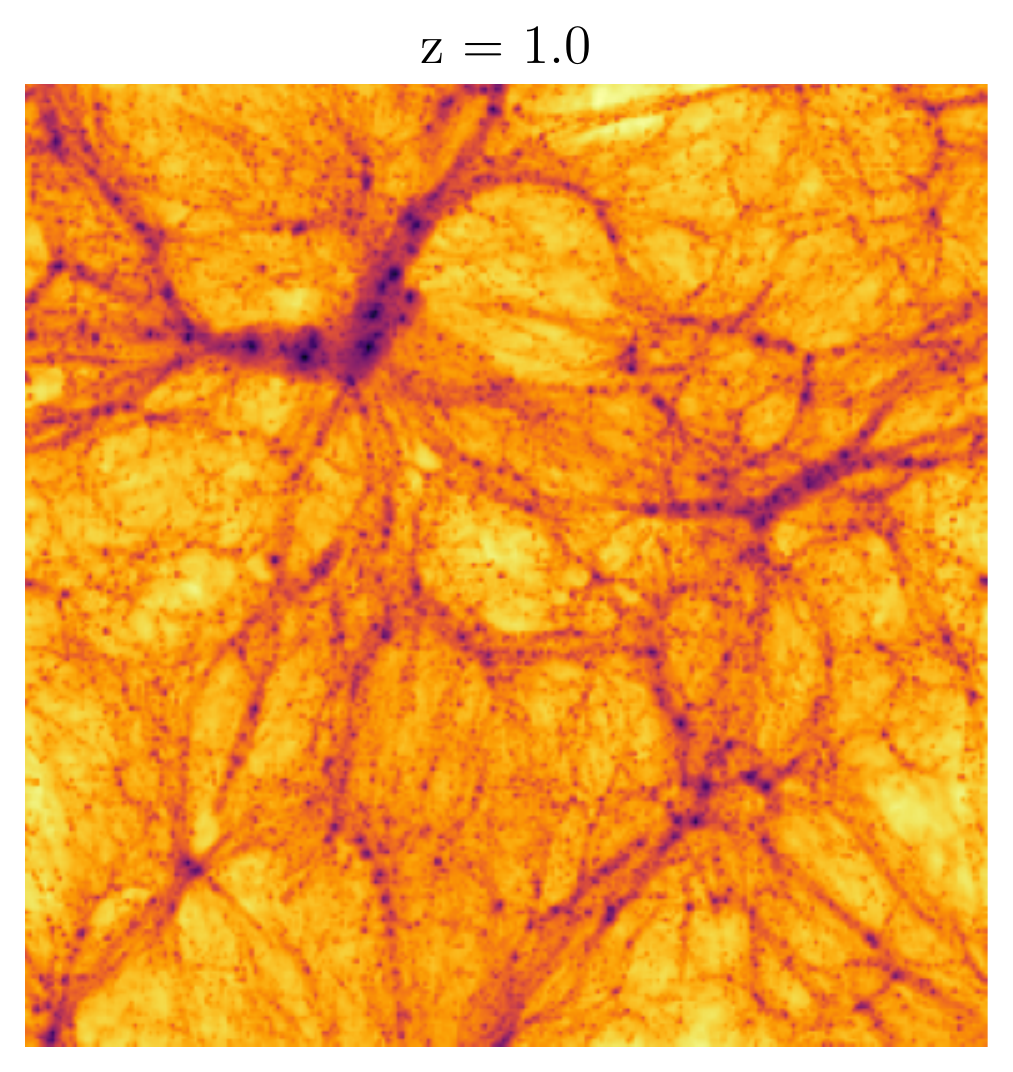}
       \includegraphics[width=.33\textwidth]{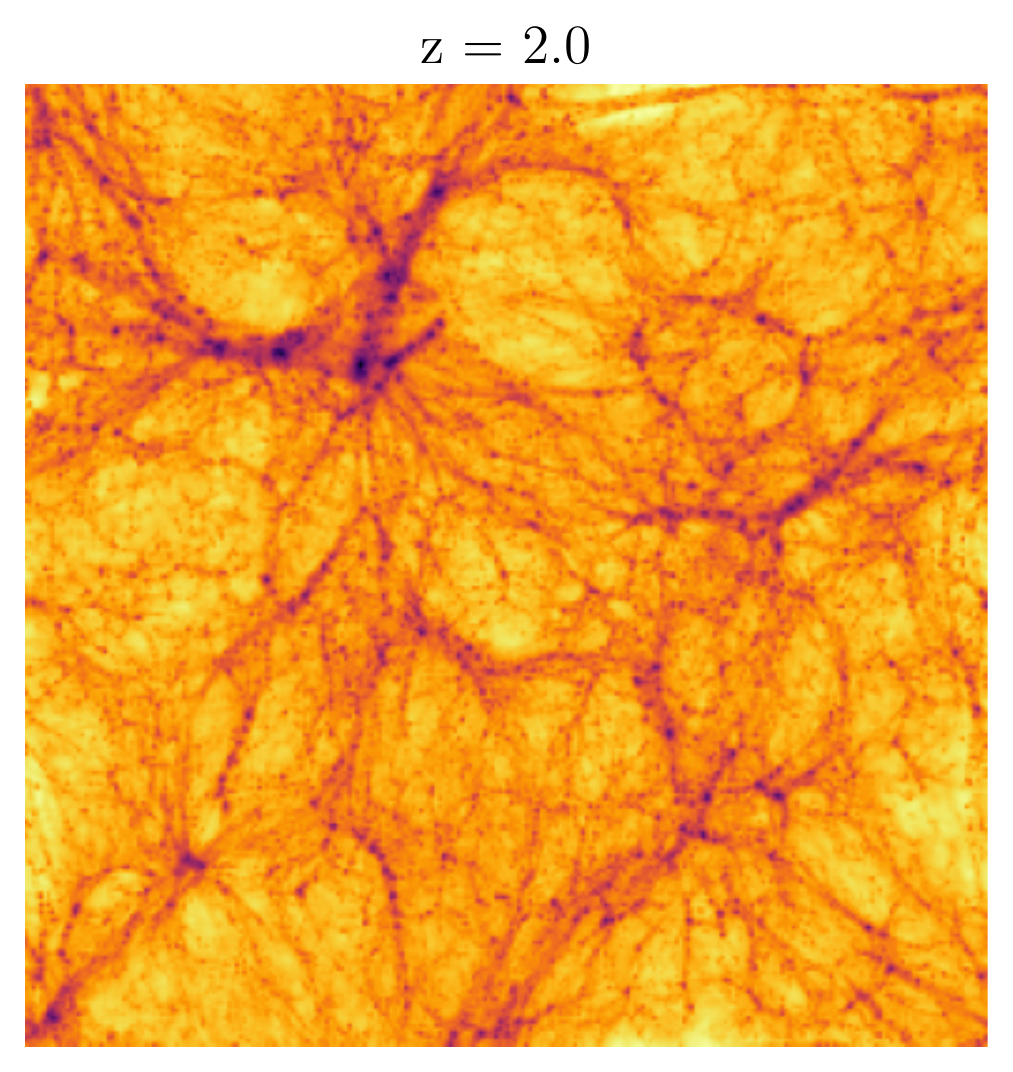}
       \includegraphics[width=.33\textwidth]{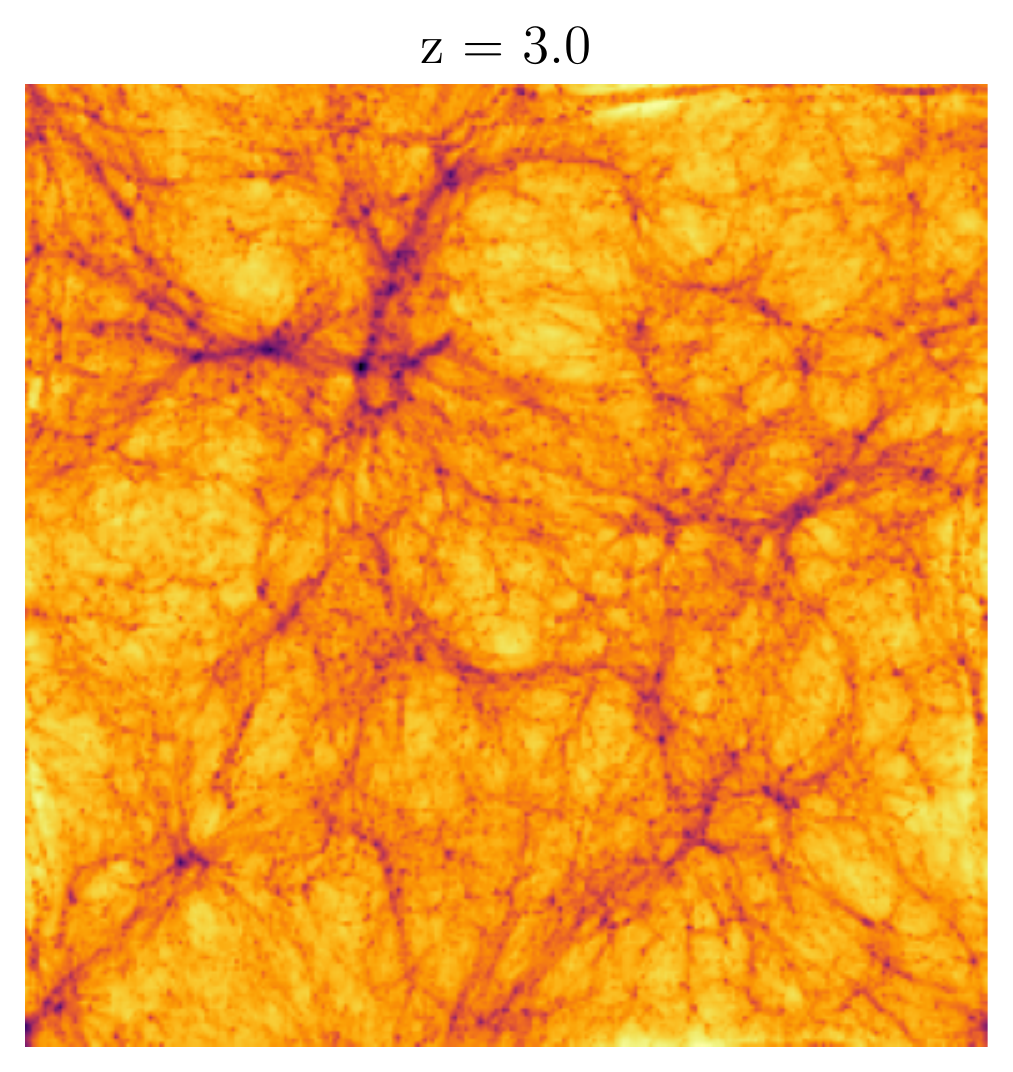}
       \includegraphics[width=.33\textwidth]{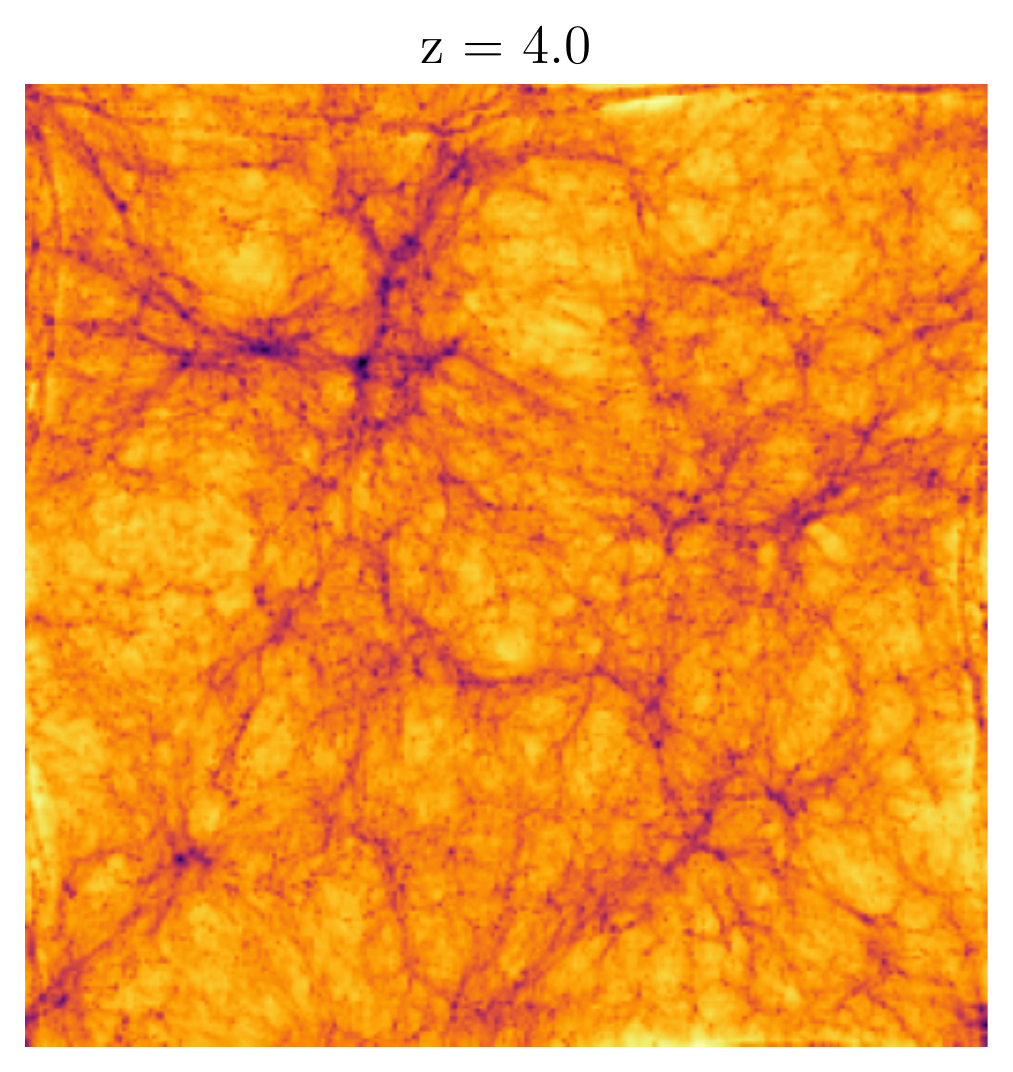}
       \includegraphics[width=.33\textwidth]{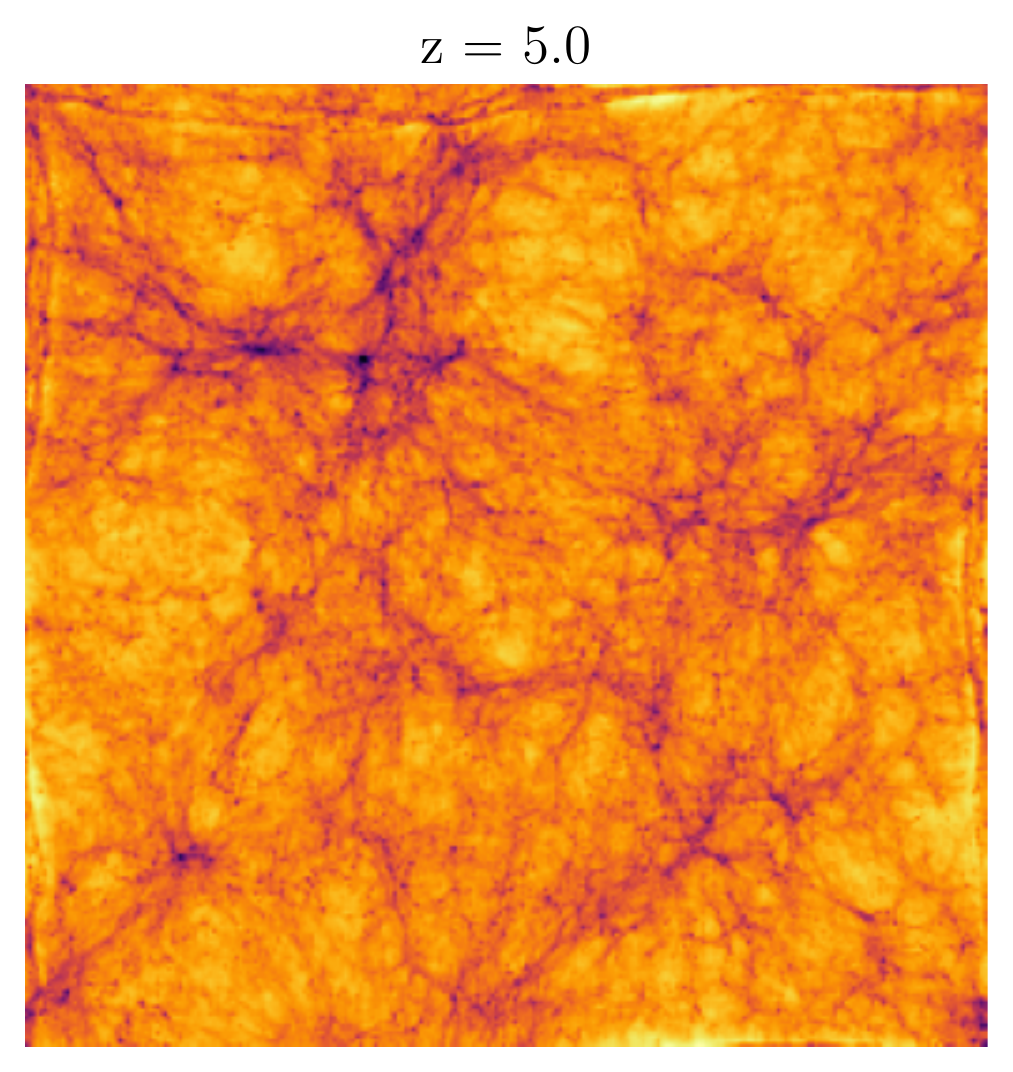}
   \caption{A comparison of images created at the different redshifts we test in Section \ref{sec:redshift}.
   Simulation and image resolutions are fixed at 512$^3$ particles and 256$^2$ pixels respectively.
   The z=0 image is the same as the image middle image in the top section of Figure \ref{fig:images}.
   }
   \label{fig:images_redshift}
\end{figure*}

Figure \ref{fig:redshift} shows how the predictive power of our models, and indirectly the amount of information that is contained in the images, changes with redshift.
Each model is trained and tuned independently with a unique set of images for the specific simulation resolution and redshift of that model.
Each set of images is fixed at an image resolution of 256$^2$ pixels.
For an example image of the same region at each redshift see Figure \ref{fig:images_redshift}.

For our fiducial high-res model, we find that the model can consistently make predictions with a 95\% cl. up to $\sim$6 keV.
There is a slight increase in the predictive power at z=0 and a slight decrease at z=4.
We do not find a trend in the predictive power with redshift up to z=5.

For our medium-resolution model, we find little change in predictive power from z=0 to z=2.
From z=2 to z=4 there is a marginal increase where the limit to which we can make accurate predictions increases from 3.7 keV to 5.2 keV.
However, the predictive power decreases again to 4.0 keV at z=5.

For our low-resolution model, we find that the model cannot make any predictions at a 95\% cl. for the WDM particle masses we explore here up to z=5.
We note that our analysis does not include a high sampling in redshift space so it may be unwise to attribute significant meaning to the trends present here.
Given the lack of a systematic trend in our results, we suspect that different predictive power across redshift is more dependent on our definition of our best accurate model than indicative of the information present in the data.
However, we discuss this data further in Section \ref{sec:saliency}.

\subsection{Varied Cosmology}
\label{sec:variedCosmo}

\begin{figure}
	\includegraphics[width=\columnwidth]{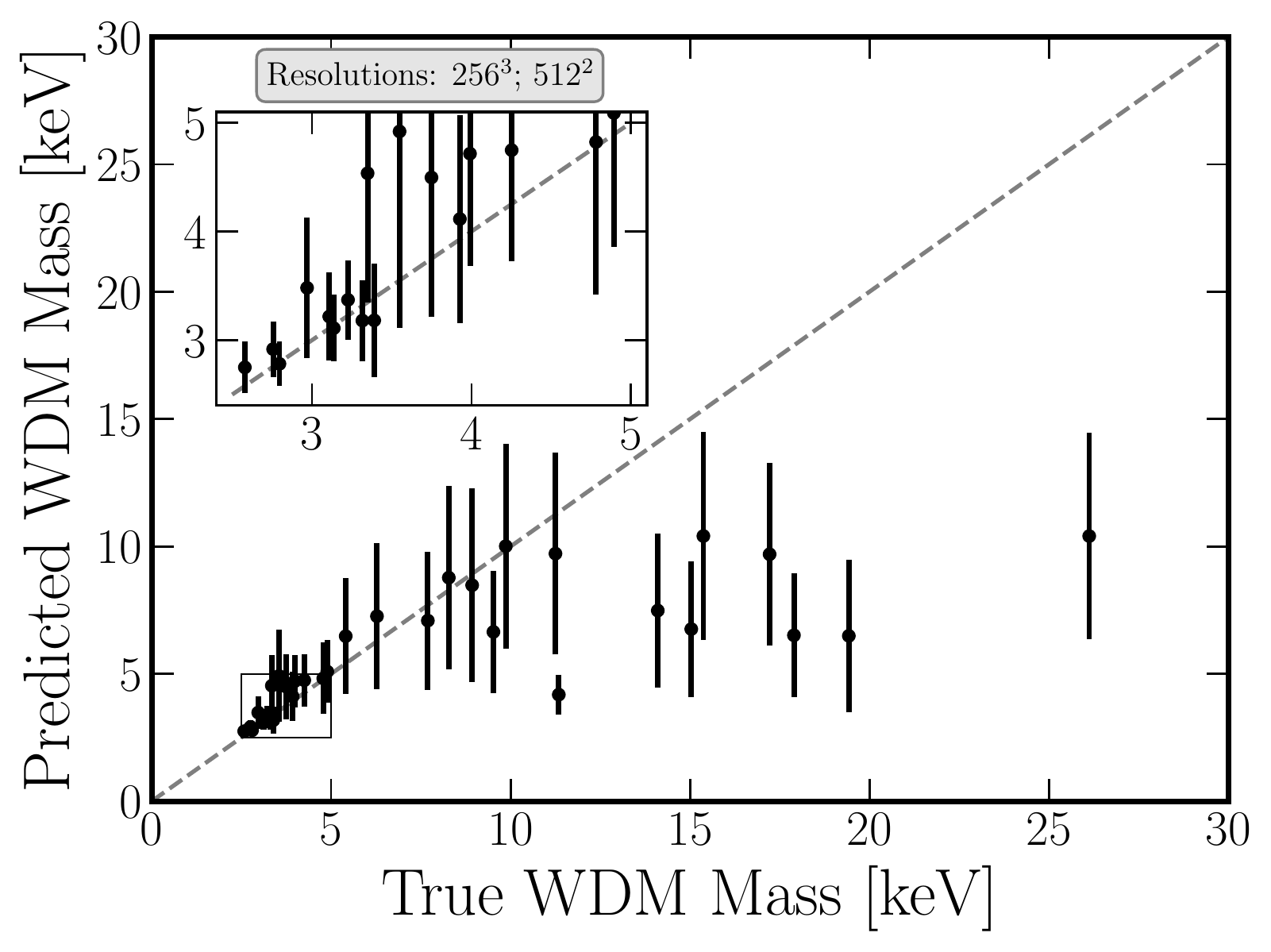}
    \caption{Similar to Figure \ref{fig:results_fixed}, but for the varied cosmology simulation suite.
    This model can make accurate predictions up to 3.5 keV with an average uncertainty of 0.6 keV and an average error of 0.5 keV.
    From 3.5 keV to 5 keV, this model can still make accurate, but less precise predictions, with an average uncertainty of 1.2 keV and an average error of 0.6 keV.
    After 5 keV, the model cannot distinguish between different WDM models and predicts an average value of $\sim$8 keV.}
    \label{fig:results_varied}
\end{figure}

Figure \ref{fig:results_varied} shows the results for our simulation suite which varies cosmology as well as the WDM particle mass.
This model has been independently trained and tested from the models presented previously.
The simulations that are used to train and test the model are at the same simulation and image resolutions as our medium resolution model (256$^3$ particles; 256$^2$ pixels). 

This model can accurately make predictions up to 3.5 keV with an average uncertainty of 0.6 keV and an average error of 0.5 keV.
From 3.5 keV to 5 keV, the model can still make accurate, although more uncertain predictions, with an average uncertainty of 1.2 keV and an average error of 0.6 keV.
Similar to the fixed cosmology model, this model reaches a point (near 5 keV) where it guesses an average value once it can no longer distinguish between different WDM particle masses.
Overall, this model does not perform as well as the fixed cosmology model at all WDM particle masses tested here.

Similar to our models presented in Figures \ref{fig:im_res} and \ref{fig:sim_res}, these results include some points which fall below the expected value.
A possible explanation for these outliers is that features in these images are degenerate with warmer WDM models.
We will discuss these outliers more in Section \ref{sec:outliers}.

\section{Discussion}
\label{sec:discussion}

In the previous section, we present results from various models trained and tested on datasets that vary in image resolution, simulation resolution, redshift, and cosmology.
We found that each of these variations affected the predictive power of our models.
In this section, we discuss what these variations tell us about where the discerning information is contained at the field level which can differentiate between different WDM models.

In Section \ref{sec:powerSpec} we compare our results from our fiducial model with a model trained solely on the power spectrum created from the same data.
In Section \ref{sec:saliency} we discuss the results from the model variations presented in the previous section.
In Section \ref{sec:fragmentation} we investigate whether numerical artifacts, such as artificial fragmentation heavily influence our results.
In Section \ref{sec:outliers} we discuss the presence of outliers in our results and what these may tell us about how our models make their predictions.
Finally, in Section \ref{sec:baryons} we discuss this investigation's limitations and possible future augmentations.

\subsection{Field-level versus Power Spectrum}
\label{sec:powerSpec}

In this paper, we have introduced a new method to infer the WDM mass directly from 2D total matter density maps that employ CNNs. We now ask ourselves whether our method yields tighter constraints than traditional methods, like the power spectrum. Here, we have focused our attention on images generated from the high resolution (512$^3$ particles) fixed cosmology suite that has $512^2$ pixels, but note that our conclusions are the same if we use other setups.

\begin{figure}
 	\includegraphics[width=\columnwidth]{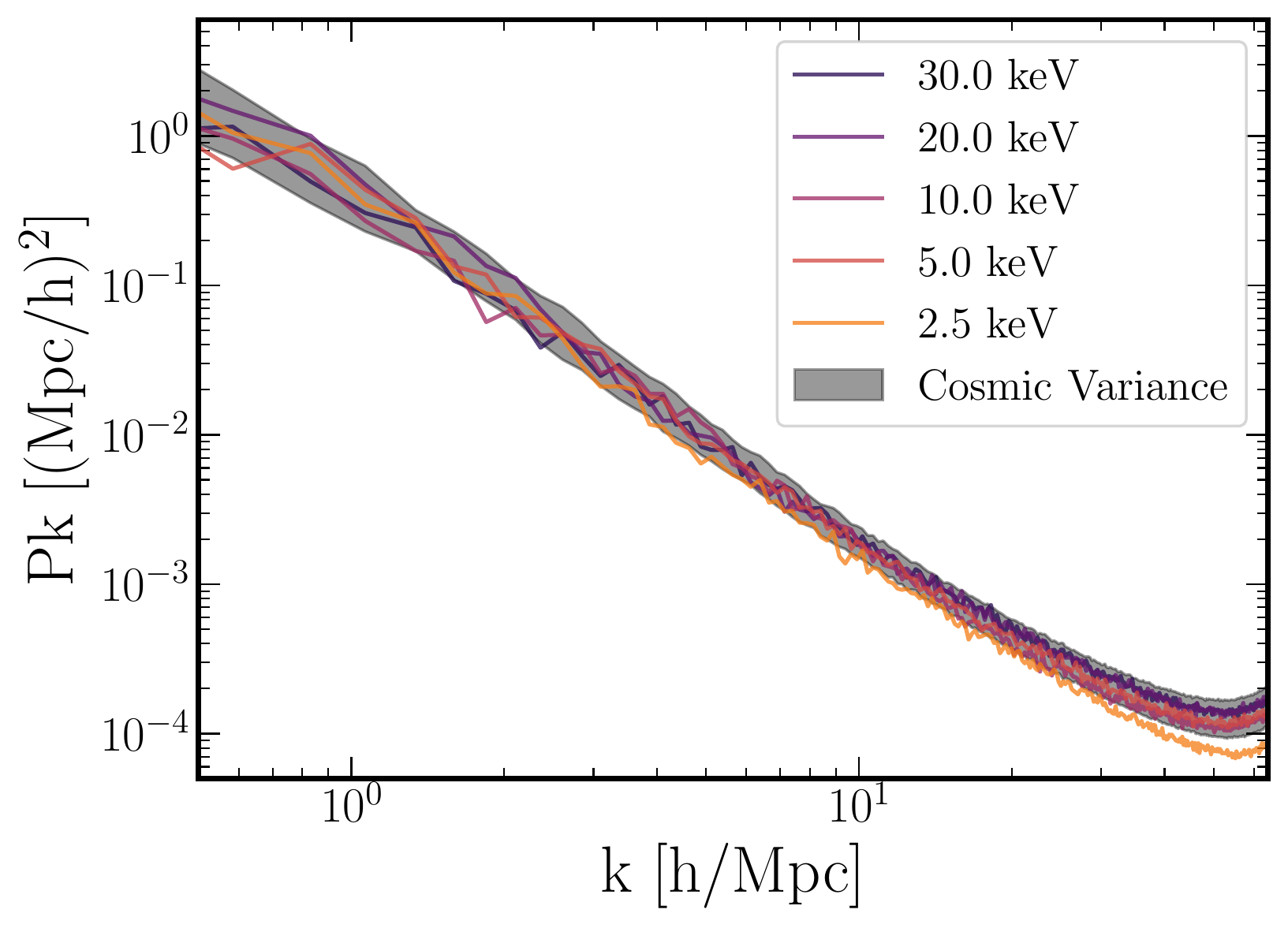}
    \caption{A sample of 2D power spectra measured from our fiducial image set for a range of WDM particle masses. 
    At large scales (small wavenumber), cosmic variance dominates the differences between the power spectra. 
    At small scales (high wavenumber), WDM suppression can be seen for the most extreme models.
    This figure also includes the expected 2$\sigma$ spread in the power spectra from cosmic variance.
    This spread is measured from all images in our five coldest training simulations which are indistinguishable from CDM so all variance is due to differences in the initial random matter distribution.
    The spectra are measured from the same images that are used to train and test our fiducial model.
    }
    \label{fig:power_spec}
\end{figure}

In Figure \ref{fig:power_spec}, we show power spectra from different images with different WDM masses: 2.5, 5, 10, 20, and 30 keV. We have computed the power spectrum of these images using the Pylians library\footnote{https://pylians3.readthedocs.io} \citep{Pylians}. We note that the different power spectra arise from maps from simulations run with different seeds, so there is not a simple monotonic relation between the amplitude of the power spectrum and the WDM mass. 

This does not exactly match current methods that use power spectra to differentiate between different physics models.
Instead, we choose this method to provide the most even-handed comparison with the field-level analysis we present here.
Current methods which compare power spectra often cover larger areas than our 25 h$^{-1}$ Mpc boxes \citep{2019DeRose,2021Angulo} or employ variance-reducing techniques \citep{2016Angulo, 2022Kokron} to not be as affected by cosmic variance.
These techniques improve the effectiveness of a power-spectra-only analysis from the results we present here where we use a small volume with a varied random seed.

However, similar variance-reducing techniques could also be applied to our field-level analysis to achieve better results.
For example, a field-level analysis on a larger simulation volume with fixed random seeds \citep[eg.][]{2018Garrison} would likely produce better results than our fiducial model.
However, producing enough of these large simulation suites to produce our training set at the desired resolution is too computationally expensive.
Varying the random seed also allows our models to produce results that are insensitive to the exact cosmic structure present in our simulations.
Therefore, the power spectra results we present here may not reflect the maximal constraining power that can be achieved from this kind of analysis, but our method does provide an even-handed comparison with our field-level analysis which could be improved by similar changes.

Figure \ref{fig:power_spec} also shows the $2\sigma$ spread in the power spectra from all images taken from the 5 coldest simulations we include in our training set.
These simulations are cold enough that the WDM particle mass is indistinguishable from CDM at this simulation resolution so any differences in the power spectrum are due to differences in the specific initial random density field that the simulations started from.
We refer to this spread in the power spectra from different random seeds as the cosmic variance.
This variance can be degenerate with effects from WDM suppression so we include the expected spread to illustrate where the WDM suppression may not be detectable over changes from cosmic variance.

As can be seen, models with low WDM masses exhibit a lack of power on very small scales, a signature of the suppression of power on these scales in the initial conditions. This is the traditional signature used to constrain WDM mass. We now quantify how accurately we can infer that parameter from the power spectra. 

We perform this task as follows. First, we compute the power spectrum of all maps. Next, we split the data into training, validation, and testing sets in the same way as outlined in Section \ref{sec:TrainingValidation} for our field-level images. The architecture of our model consists of a series of fully connected layers with LeakyReLU activation function and dropout. The loss is the same as Eq. \ref{eq:loss} and chosen such as the networks predict the posterior mean and standard deviation of the WDM particle mass.

We also train this model using \textsc{optuna} \citep{optuna} over 50 trials, similar to our field-level models.
The hyperparameters that are tuned for this model are the number of fully connected layers (NL), the learning rate (LR), and the weight decay (WD).
For each layer, there are then two additional hyperparameters that are optimized: the number of units (NU) and the dropout rate (DR).
We list the ranges over which we optimize these values and the optimal values we find in Table \ref{tab:power_spec}.

\begin{table}
    \centering
    \begin{tabular}{l|lll}
    \hline 
        Name & Min  & Max  & Optimized Value  \\
        \hline 
        NL   & 1    & 5    & 3 \\
        LR   & 1e-5 & 1e-1 & 1.5e-4 \\
        WD   & 1e-8 & 1e0  & 5.9e-2 \\
        DR   & 0.2  & 0.8  & (0.73, 0.35, 0.80) \\
        NU   & 4    & 1000 & (217, 612, 358) \\
        \hline 
    \end{tabular}
    \caption{List of hyperparameters for our model used on the power spectrum created from the images used in our fiducial models. 
    Column one lists the names of the parameters: number of neuron layers (NL), learning rate (LR), weight decay (WD), dropout rate (WD), and number of units (NU).
    The second and third columns list the minimum and maximum values we tune our models over.
    We use \textsc{optuna} to train our models over 50 trails where each parameter is varied simultaneously.
    The optimal trial is chosen as the one with the lowest validation loss, and the optimal values are shown in column four.
    The last two parameters, DR and NU, are used in each neuron layer. 
    Since this model contains three neuron layers, the optimized value for each layer is shared.
    The results for this model are shown in Figure \ref{fig:power_spec_results}.
    For a discussion on this model, see Section \ref{sec:powerSpec}. }
    \label{tab:power_spec}
\end{table}

\begin{figure}
	\includegraphics[width=\columnwidth]{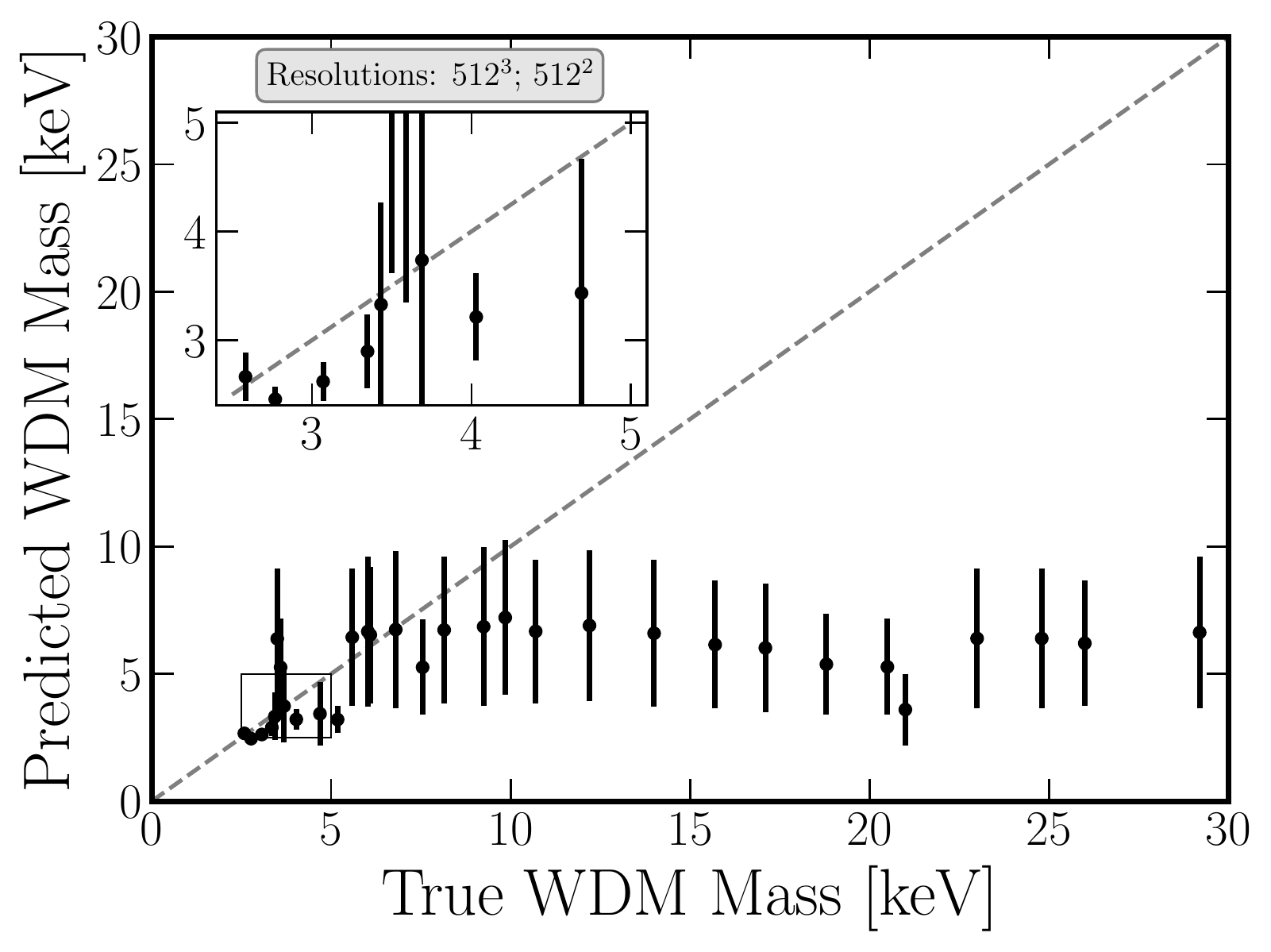}
    \caption{Results from the model trained on the power spectra shown in Figure \ref{fig:power_spec}. This model can make accurate predictions up to 5.5 keV with an average uncertainty of 0.60 keV and an average error of 1.0 keV. See text in Section \ref{sec:powerSpec} for a more detailed discussion.}
    \label{fig:power_spec_results}
\end{figure}

In Figure \ref{fig:power_spec_results} we present our results from the model trained on the power spectra.
We find that the model can make accurate predictions on the WDM particle mass from 2.5 keV to 5 keV.
In this range, the model has an average uncertainty of 0.6 keV and an average error of 1.0 keV.
Between 4.0 and 5.5 keV, the model underpredicts the correct WDM particle mass and appears to guess an average value of $\sim$3 keV.
Likely, the model can distinguish these spectra from colder spectra, but cannot differentiate between them due to interference from cosmic variance.

There are two outliers near 3.5 keV which are predicted to be much higher than the true value.
It appears that other images taken from the same simulation produce values which match the correct value better than the ones randomly chosen here.
Most likely, the specific cosmic variance in the field which created those power spectra obscured the suppression from WDM.

Our fiducial model is trained with the same images used to create these power spectra. 
Therefore, our fiducial results presented in Figure \ref{fig:results_fixed} can be compared to our power spectra results shown in Figure \ref{fig:power_spec_results}.
Over the same section of WDM parameter space that the power spectra model can make predictions (2.5-5.5 keV), our fiducial model can make predictions with an average uncertainty of 0.82 keV and an average error of 0.52 keV.
This represents a $\sim$2$\times$ increase in precision in this WDM mass range from the model trained on power spectra presented here.
Additionally, our fiducial model can make accurate predictions up to 10 keV, whereas this model can only make (less precise) predictions to 5.5 keV.

Compared to the model that works on images  (see Figure \ref{fig:results_fixed}), we find a general worsening.  As expected, extracting information from the field itself should yield more accurate results, as we find. 
From this, it is likely that
the usage of summary statistics implies the loss of information. 

\cite{2021Villaescusa} did a similar comparison of field-level to power spectrum inference for cosmological parameters.
Similarly, they find that their CNN models extract information beyond just the power spectrum when field-level data is available.
Further, they show that their models extract information from field-level data at many size scales available.
For WDM, the power suppression only appears in the matter power spectrum of the initial conditions, non-linear gravitational evolution will move information from two-point correlations to higher-order moments. 
Thus, it is expected that the power spectrum does not contain all available information concerning the WDM mass.

\subsection{Model Interpretability}
\label{sec:saliency}

%



In this section, we attempt to demystify the inner workings of the models we present to gain insight into the physical processes that can be used to differentiate between different WDM models.
CNNs are notoriously black boxes that do not give insights into how they arrive at their predictions.
We discuss what the results from the model variations and power spectra models tell us about how these models make their predictions.

One way to parameterize the amount of suppression from WDM that the CNNs can detect is to measure the suppression in the simulation's initial conditions at the Nyquist frequency.
Nonlinear growth will either wipe out or move suppression at higher frequencies where the suppression is strongest, so measuring the suppression at the initial conditions can provide a measure of the maximal amount of signal that may be present at later times.
This can help us to place an upper limit on where the model can be expected to distinguish between different models as the maximum amount of information in the image may be very small.

In Figure \ref{fig:resolution}, we find that the predictive power of our models relies heavily on the simulation resolution and minimally on the image resolution.
For our high- (medium-) simulation-resolution model, the Nyquist frequency is at 64.3 (32.2) h Mpc$^{-1}$, and the model can make predictions with at least 95\% confidence up to 7.5 (4.7) keV.
This corresponds to 2-9\% (1-6\%) suppression from CDM between 0.25x and 0.5x the Nyquist frequency.
Our low-resolution model has a Nyquist frequency of 16.1  h Mpc$^{-1}$, but the model cannot make any predictions with at least 95\% confidence for any of the WDM masses that we test.
For our warmest WDM mass, 2.5 keV, this corresponds to 1-7\% suppression between 0.25x and 0.5x the Nyquist frequency, comparable with our other resolutions.
While our results are not completely consistent and likely do not capture the complexities of how the model measures this suppression at late times, they suggest that these models can make accurate predictions from $\sim$5\% reduction in power at 0.5x the Nyquist frequency in the initial conditions.

We find that the Nyquist frequency of the images does not strongly correlate with our models' predictive power, even if it is significantly smaller than the Nyquist frequency of the simulation particles.
This is unsurprising as the simulations at late times are no longer representative of a grid of particles with a well-defined Nyquist frequency.
The simulations' nonlinear growth most likely moves suppression that was present at high frequencies to lower frequencies which can be measured at the image resolutions we explore here.

Our inability to make any accurate predictions at our lowest resolution simulations places some constraints on where the important information is located in our data.
We know that either small size scales or low densities are required to differentiate between our WDM models.
For our lowest resolution simulation suite, the length at which numerical gravitational softening affects our results is 2.4 kpc at z=0.
Similarly, this simulation suite has a mass resolution of $6.24 \times 10^8$ $h^{-1} M_\odot$ where any smaller structures will not be present.
Therefore, there must be features present that are smaller than 2.4 kpc or $6.24 \times 10^8$ $h^{-1} M_\odot$ that our models are using to differentiate between these WDM models.
In Figure \ref{fig:resolution} we also noted the large increase in predictive power when we include at least 256$^2$ pixels for our highest resolution simulation suite.
Since the images are fixed at 25 Mpc$^2$, an image with 256$^2$ pixels will have a spatial resolution of 98 h$^{-1}$ kpc and an image with 128$^2$ pixels will have a spatial resolution of 195 h$^{-1}$ kpc.
Therefore, for this dataset, there is a strong discriminator present in the data with a linear size between 100 and 200 h$^{-1}$ kpc.

In Figure \ref{fig:redshift}, we find that there is no systematic trend in the predictive power of our models with redshift.
Suppression to the power spectrum from WDM is more apparent at higher redshifts due to the DM spending less time in its non-linear growth regime.
Since our models do not show a similar change in their predictive power, it is plausible that the models are not relying solely on the power spectra to make their predictions.
Alternatively, if there were a new discriminator born from the non-linear structure formation, we might expect the predictive power of our models to increase at lower redshift.
Instead, the consistent predictive power that we observe may be more of an indication that the information about the suppression of power in the initial conditions is still present at later times, but in a different form that can be obtained from field-level data.

In Figure \ref{fig:results_varied} we show the results of our model when we vary $\Omega_m$ and $\sigma_8$ along with the WDM particle mass.
We find that models trained on this simulation suite do not perform as well as those trained on the fixed-cosmology suite.
A possible explanation for this discrepancy is that there is a degenerate effect from these cosmology variations with WDM particle mass.
If this is the case, further exploration may reveal additional avenues to understand the effects of WDM power suppression.
Another possibility is that the loss in predictive power is due to an increase in noise from cosmic variance.
As discussed previously, an advantage to using field-level data over the power spectrum alone may be an increased ability to disentangle effects from cosmic variance. 
It is possible that the increased variance from the varied cosmology is too large for these models to disentangle.

To test if our models rely on the power spectrum to make their predictions, we compare our results to those of a model trained with the power spectrum alone.
We measure our power spectra from the same images we use to train and test our fiducial model to make an even-handed comparison.
In Figure \ref{fig:power_spec_results} we show the results of the model trained with only the power spectra, and in Figure \ref{fig:results_fixed} we show the model trained on the field-level data.
We find that the field-level data provides much more information that the model can use to predict WDM particle masses than is found in the power spectrum alone.
As was hinted at by the redshift dependence of our results, it is possible that information that was present in the initial condition's power spectrum has moved during nonlinear structure growth and is no longer captured by the power spectrum.

On the other hand, the main benefit of field-level analysis may be an increased ability to undo the effects of cosmic variance over using summary statistics alone.
There have been techniques developed to get a better insight into how these models are making their predictions, such as saliency maps, integrated gradients, and SHAP values.
Obtaining results from these methods is trivial, but incorporating them into a working understanding of the physical features that drive these models to make their predictions is not.
Therefore, we do not explore these methods here, but instead, leave this work to a future investigation focused on understanding the inner workings of these black-box models.

\subsection{Numerical Fragmentation}
\label{sec:fragmentation}

In the previous section, we discussed possible physical interpretations of our results.
As our data is taken from numerical simulations and is thus non-physical, we discuss how numerical artifacts may affect our results.
Numerical fragmentation is a particular numerical artifact inherent to simulations with low power on small scales, like the case of WDM. This phenomenon has been widely discussed and various techniques have been put forward to eliminate these from halo and galaxy catalogs created from WDM simulations.
In this section, we will argue that our simulations are not heavily affected by numerical fragmentation and thus our results are robust to these numerical systematics.

Numerical fragmentation occurs in discrete numerical simulations near the cutoff scale in WDM simulations.
When filaments form during normal cosmic structure formation, artificial power at small scales is present from the discrete sampling of the underlying density distribution.
This power then grows to form artificial halos at later times.
\cite{2007Wang} have shown that increasing the resolution of the simulation can alleviate, although not eliminate, this problem.
For our warmest model (2.5keV), the cutoff in the power spectra occurs at a wavenumber of 18 (h/Mpc) which corresponds to a half-mode mass of 9.67 $\times$ 10$^8$ M$_\odot$, or 68 (8) particles for our high (medium) resolution simulation suite \citep{2021Enzi}.
Since this cutoff occurs above the mass resolution of our simulation we can expect some artificial clumping in our warmest simulation models.
For our medium and low-resolution models, and in models with higher WDM particle masses, this effect will be less pronounced.

\begin{figure}
	\includegraphics[width=\columnwidth]{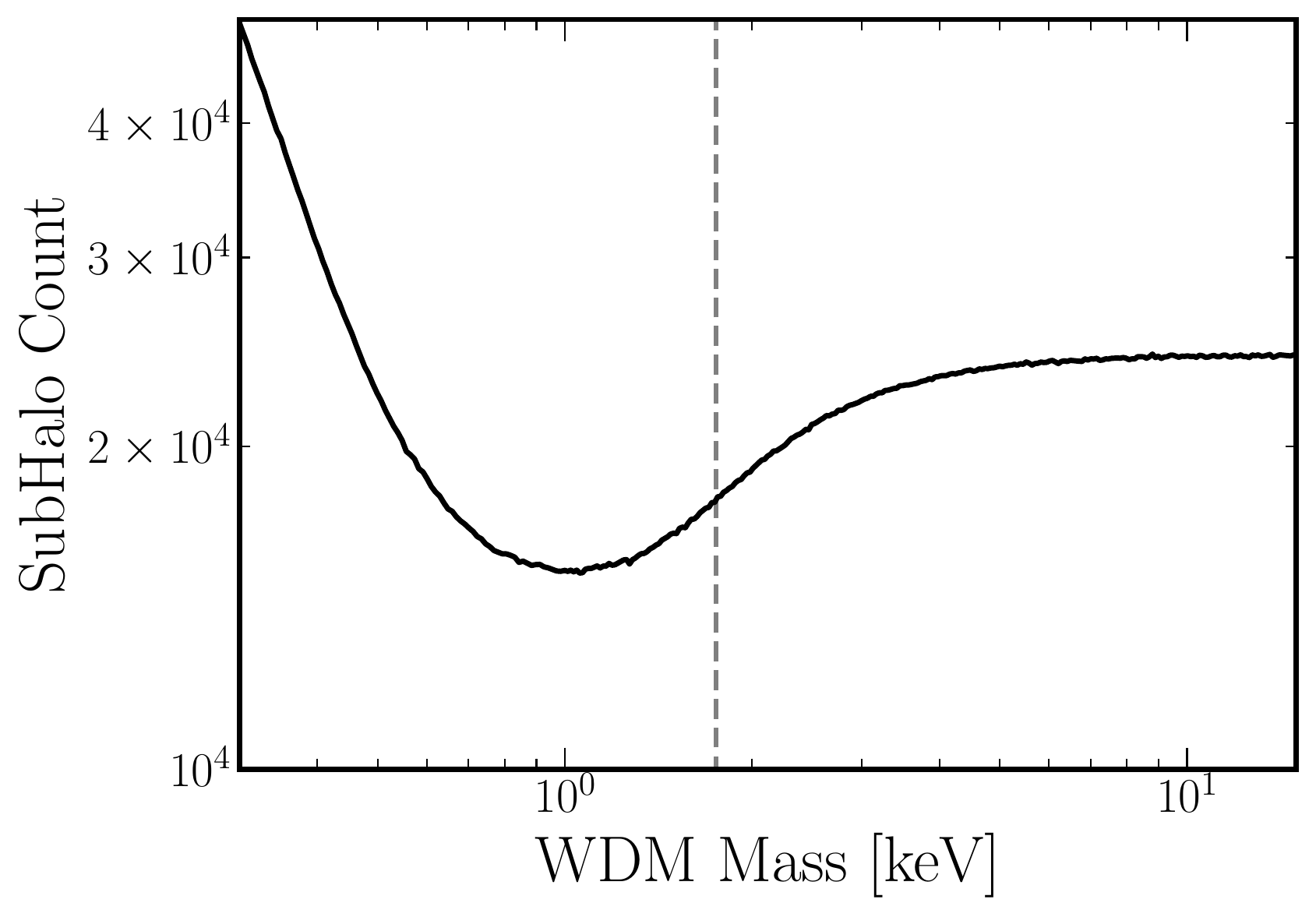}
    \caption{A comparison of the number of subhalos in a simulation vs the WDM particle mass.
    Each data point presented here represents the total number of halos in a unique simulation.
    We extend the WDM particle mass range we explore significantly to better show how and when numerical artifacts affect these simulations.
    We include a dashed line at 1.8 keV where the decrease in the number of subhalos from WDM suppression begins to decrease due to numerical fragmentation.
    Since this inflection point occurs outside the range of WDM particle masses we explore in this paper, we do not believe numerical fragmentation has a significant effect on the results we present here.
    }
    \label{fig:fragmentation}
\end{figure}

In Figure \ref{fig:fragmentation}, we show the number of DM halos, calculated with \textsc{subfind} \citep{2001Springel}, as a function of WDM particle mass.
Each datapoint shown represents a separate simulation (done at a resolution of 256$^3$ particles) with a fixed random seed and different WDM particle mass.
Since the random seed used to initialize the simulations is held constant, any decrease in the total number of halos can be attributed to suppression from WDM.
Any increase in the number of halos can be attributed to numerical fragmentation.

To emphasize the effects of numerical fragmentation, we extend the range of WDM particle masses from our main study significantly to include more extreme models.
The range of WDM particle masses presented here ranges from 0.3 to 15 keV.
At high WDM masses, where the models most similar to CDM are present, there is little deviation between the different models.
Near 3 keV, the number of subhalos begins to fall due to the now significant suppression in power from WDM.
This trend continues until 1 keV where the number of subhalos begins to increase with lighter WDM particle masses down to 0.3 keV.

To aid the reader, we have marked (dashed line) the inflection point at 1.8 keV where the number of subhalos begins to decrease at a slower rate with warmer models.
This marks where numerical fragmentation begins to counteract the suppression from WDM.
Since the warmest model that we present in this paper (2.5 keV) falls well above this inflection point (1.8 keV), we believe that numerical fragmentation does not significantly affect the results we present.

We note that this investigation was only carried out at one simulation resolution with a fixed cosmic seed.
\cite{2007Wang} has shown that simulation resolution can affect the abundance of numerical fragmentation so the results we present in Figure \ref{fig:fragmentation} may not hold for all resolutions.
Additionally, different areas within the cosmic structure may be more affected by numerical fragmentation than others.
Since our main results are gathered from simulations with varied initial random seeds, it is possible that for a different simulation, at the same resolution, the results presented in Figure \ref{fig:fragmentation} may differ.

However, the driving force behind the numerical fragmentation is the cutoff frequency of the WDM power suppression. 
Since the range of WDM masses, we explore is fixed, we do not expect numerical fragmentation to play a significant role in any of the results we present here.

\subsection{Outliers}
\label{sec:outliers}

Within the results we present here, there are some outliers present that warrant further discussion.
The first of these is the systematic offset at high WDM particle masses between the predicted and true values that is present in all of our models.
The second is individual points which are predicted to be far from their true value, often with small errorbars.

Toward the first point, the systematic offset between the predicted and true values at high particle masses occurs once a model is no longer able to distinguish between different WDM models.
This transition occurs in each model we present, although the specific particle mass at which this occurs changes for each model depending on its predictive power.
Likewise, the average value that the model predicts also changes with each model depending on where the transition occurs.

The uncertainties on these points given by the model are often under-represented in what they should be to account for their distance from the true value.
This discrepancy can be attributed to the formulation of our loss function (see Equation \ref{eq:loss}).
According to our loss function, there is a set error that is attributed to each point depending on what the model predicts the WDM mass to be.
Therefore, if the model believes that an image has a WDM particle mass of 5keV, it will prescribe the same error to it regardless of whether the image is truly at 5keV.
Since the model does not know the true value while making its prediction, this is not an easy shortcoming to surmount.

One might expect that the model would guess an average value between the model where it can make an accurate prediction and the maximum particle mass that we explore to minimize its loss during training.
If we assume for the moment this is true, we can explore whether this can be used as an accurate indicator of the extent to which our model can make accurate predictions.
For our fiducial model, the average value that it predicts for values greater than 10 keV is 9.5 keV.
We exclude the outlier near 17keV in this calculation.
Accounting for the non-uniform distribution of simulations upon which the model is tested, this would provide us with a maximum confident prediction of 5.8 keV.

We show in Section \ref{sec:variations} that we can distinguish accurate predictions from the points that lay after the transition to `guesses' up to 7.5keV at a 95\% confidence level.
Since this is a much greater value than the one we just calculated, it appears the model does not predict an average value of the models it cannot differentiate between.
A more likely explanation may be that the model predicts just above the maximum mass it can accurately predict to ensure some predictions are correct.
While this does not affect any results we present here, a better understanding of what these models do when they are uncertain of the result can allow us to better interpret our results in the future.

The second type of outlier we discuss here is points that occur after the transition to `guesses' (high WDM particle masses), but which are incorrectly predicted to be at a lower mass.
These types of outliers are present in most of the results we present in Section \ref{sec:results} to some degree and are more common in lower-resolution models.
To understand whether these points are incorrect due to a random error or something intrinsic to the data, we retrain five additional independent CNNs with the same tuned hyperparameters as the original model.
For nearly all of these outliers, each independently trained model results in a similar inaccurate prediction.

Even with both systematic and random errors included, these data points still fall many standard deviations away from the true value, or the expected average guess for high-mass models.
These points represent areas of the simulations that can repeatedly fool the neural networks into thinking that their DM properties are different from their true nature.
While a full investigation into these outliers is not included in this paper, they may include further insight into indicators that can be used to differentiate between different DM models.

\subsection{Baryons}
\label{sec:baryons}

One shortcoming of our investigation presented here is the lack of baryons and a comprehensive galaxy formation model in our simulations.
Many authors have shown that including baryons in simulations, with both hydrodynamics and a comprehensive galaxy formation formulation, greatly increases the accuracy of the simulation concerning observations \citep[for a review see][]{2020Vogelsberger}.
Baryons themselves have a significant influence on the properties of the DM halos they populate, and their specific prescription can greatly vary the final halo state \citep{2010Tissera, 2020Callingham}.

For alternative DM models, the effects from baryons can be degenerate with effects from the specific model and can overpower any signal that summary statistics can detect \citep{2011Semboloni, 2022Moreno, 2022Mastromarino, 2023Rose}.
While this is not as prevalent for WDM, this may reduce the predictive power of our models, even if we include additional baryonic properties.
In \cite{CAMELS}, the authors showed that CNN models with the same architecture that we present here can make predictions of both baryonic properties and cosmology.
This may hint that these models may be able to differentiate the effects of baryons from alternative DM models better than we can with our current summary statistics.
Furthermore, the inclusion of baryons may improve the accuracy and precision of the results presented here.
By including additional baryonic projections along with the DM density images presented here, the models may be able to break some degeneracies that caused the outliers discussed in the previous section.
However, given that the changes induced by alternative DM models are often overpowered by effects from baryons, a full investigation into how these models can disentangle and utilize baryonic signals is needed.

\section{Conclusions}
\label{sec:conclusion}

We presented the first investigation in utilizing neural networks to differentiate between different DM models through a field-level approach with WDM.
We explore many model variations, including simulation resolution, image resolution, cosmology, and redshift to understand the information pertinent to different WDM models.
We also compare our results to those derived solely from power spectra data to better understand what information is gained at the field level.
Our main conclusions are summarized below:

\begin{enumerate}
    \item We can train a neural network to accurately predict WDM particle masses up to $\sim$10 keV, 
    with an uncertainty of $\pm$ 1keV,
    in our fiducial setup. See Section \ref{sec:fixedCosmo}.
    \item Increases in simulation resolution have the strongest effect on the predictive power of our models. The predictive power also increases slightly with increased image resolution. For our high-resolution model, we find an important feature exists with a linear size between 100 and 200 h$^{-1}$ kpc. We find no systematic increase in predictive power with redshift up to z=5. See Section \ref{sec:evaluation}.
    \item Accurate predictions can be made across many redshifts, simulation resolutions, and image resolutions with at least a 95\% confidence level. See Section \ref{sec:evaluation}.
    \item Predictions of WDM particle masses up to 5.5 keV remain accurate, even when varying $\Omega_m$ and $\sigma_8$, but with increased uncertainty and error. See Section \ref{sec:variedCosmo}.
    \item Field-level inference can make better predictions than power spectrum analysis suggesting that additional information than just the power spectrum can be used to better constrain DM models. See Section \ref{sec:powerSpec}.
    \item The model is learning physical features, not numerical artifacts. See Section \ref{sec:fragmentation}.
\end{enumerate}

We find that our models can accurately differentiate different DM models across a range of parameter space that is not easily probed by other methods.
For our fiducial model, we can make accurate predictions up to 7.5 keV at a 95\% confidence level.
The model can also make accurate predictions up to 10 keV, but the predictions become indistinguishable from higher particle mass predictions.

We also find that our model is flexible enough to make accurate predictions with different simulation resolutions, image resolutions, redshifts, and cosmologies.
We find minimal dependence on image resolution, although there may be a simulation resolution-dependant size scale that must be resolved before additional resolution does not affect the results.
We find that simulation resolution has the largest effect on the predictive power of our models.
We find no systematic trend between our models' predictive power and redshift up to z=5.
Our model is still able to make an accurate prediction, although to a lesser degree once we vary the cosmology along with WDM particle mass.

Using our fiducial model and its variations, we can determine some information about how our models are making their predictions.
From the large increase in predictive power in our high-resolution model from increasing the image resolution up to 192$^2$ pixels, we can infer that DM properties around 130 kpc in size exist that can differentiate between DM models.
From the strong dependence on simulation resolution, we infer that large-scale structures alone are not sufficient to make any accurate determination of WDM properties.

By examining data at multiple redshifts, we find no systematic trend in our predictive power from z=0 to z=5.
This suggests that the information that is present in the power spectrum at high redshifts is still present, but requires full field-level data to access it.
When we compare our results to similar results derived from the same data but limited to just the power spectrum, we find that field-level data either contains additional information not present in the power spectrum or that degeneracies due to cosmic variance are easier to break at the field level.
In either case, we find that a field-level analysis provides significantly more discerning power between different WDM models than the power spectrum alone.

While a more in-depth investigation into interpreting our models' predictions is necessary, we show that we can produce a model that uses physical features to differentiate between alternative DM models.
This opens opportunities to explore more realistic DM models and simulation techniques to gain further knowledge on how we can determine the properties of the DM in our Universe.

\section*{Acknowledgements}

The authors acknowledge University of Florida Research Computing for providing computational resources and support that have contributed to the research results reported in this publication.
This work used Anvil at Purdue University through allocation PHY220051 from the Advanced Cyberinfrastructure Coordination Ecosystem: Services \& Support (ACCESS) program, which is supported by National Science Foundation grants \#2138259, \#2138286, \#2138307, \#2137603, and \#2138296.

JCR acknowledges support from the University of Florida Graduate School’s Graduate Research Fellowship.
PT acknowledges support from NSF grant AST-2008490 and NASA ATP Grant 80NSSC22K0716. 
The work of FVN is supported by NSF grant AST 2108078.
MV acknowledges support through NASA ATP 19-ATP19-0019, 19-ATP19-0020, 19-ATP19-0167, and NSF grants AST-1814053, AST-1814259, AST-1909831, AST-2007355 and AST-2107724.
MM acknowledges support through DOE EPSCOR DE-SC0019474 and NSF PHY-2010109.
DAA acknowledges support by NSF grants AST-2009687 and AST-2108944, CXO grant TM2-23006X, Simons Foundation Award CCA-1018464, and Cottrell Scholar Award CS-CSA-2023-028 by the Research Corporation for Science Advancement.

\section*{Data Availability}

The data and code used to produce this paper can be made available upon reasonable request to the corresponding author.

\bsp	
\label{lastpage}
\end{document}